\documentclass[11pt]{article}
\usepackage{geometry}                
\usepackage{graphicx}
\usepackage{amssymb}
\usepackage{mathtools}
\usepackage{epstopdf}
\usepackage{verbatim}
\usepackage{color}
\usepackage{slashed}
\usepackage{bbold}
\usepackage{fullpage}
\usepackage{authblk}
\usepackage{float}
\usepackage[title]{appendix}
\usepackage{caption}
\usepackage{subcaption}
\usepackage{cite}

\restylefloat{table}
\restylefloat{figure}

\newcommand{\beq}{\begin{equation}}
\newcommand{\eeq}{\end{equation}}
\newcommand{\bmat}{\begin{pmatrix}}
\newcommand{\emat}{\end{pmatrix}}
\newcommand{\bal}{\begin{align}}
\newcommand{\eal}{\end{align}}

\newcommand{\bm}{\boldsymbol}

\newcommand{\Order}{\mathcal{O}}
\newcommand{\sg}{\tilde{g}}

\newcommand{\kab}{\kappa_{\alpha\beta}}
\newcommand{\cab}{c_{\alpha\beta}}

\begin{document}

\title{\begin{flushright}\small{MCTP-15-07}\end{flushright} 
~\\
~~\\
\huge\bf{Lepton Flavour Violation via the K\"ahler Potential in Compactified M-Theory}}

\author{\normalsize\bf{Sebastian~A.~R.~Ellis\footnote{sarellis@umich.edu}~}}
\author{\normalsize\bf{Gordon~L.~Kane\footnote{gkane@umich.edu}~}}
\affil{\it{Michigan Center for Theoretical Physics (MCTP),}\\ \it{Department of Physics, University of Michigan}, \\ \it{Ann Arbor, MI 48109, USA}}
\date{\small{\today}}
\maketitle
\begin{abstract}
We use lepton-flavour violating (LFV) processes as a probe of higher-order corrections to the K\"ahler potential in compactified M-theory. We consider a generic K\"ahler potential with higher-order terms coupling visible sector fields to fields in the hidden sector of the compactified theory. Such terms generally give rise to potentially large flavour-violating effects. Unless there are suppressions, the size of the resulting off-diagonal terms in the K\"ahler potential may be at odds with experimental results. The rare decay $\mu \to e \gamma$ and $\mu \to e$ conversion in nuclei probe the size of the potential flavour non-diagonality of the higher-order terms for realistic spectra in the M-theory compactification. We consider a parameterisation of the higher-order corrections in terms of a small parameter $\epsilon$. By analysing various textures for the higher-order corrections, we find current bounds on $\epsilon$ from the LFV processes. The constraint from the neutral kaon mass difference $\Delta m_K$ is currently similar to that from $\mu \to e \gamma$. Measurement or new limits on the process $\mu \to e \gamma$ and, in the future, $\mu \to e$ conversion in Aluminium, will be an effective probe of the form of the higher-order K\"ahler potential terms. For the preferred range of gravitino masses, unless the K\"ahler potential is strikingly flavour-diagonal, improvement in experimental sensitivity of LFV processes should give a non-zero signal.
\end{abstract}
\newpage
\section{Introduction}
\label{Intro.SEC}

In the Standard Model (SM), lepton-flavour number is conserved because one can always rotate to a basis in which gauge interactions and the charged lepton Yukawa matrix are diagonal. Therefore, any future observation of lepton-flavour violation (LFV) would constitute evidence for new physics beyond the Standard Model (BSM). Measurement and analysis of low-energy LFV processes may provide insight on potential new physics at the high-energy frontier, since in all approaches to extending the SM, new physics will lead to LFV processes. The current limits on $\mu \to e \gamma$ and other processes present challenges for many scenarios for new physics.

In minimal supersymmetric extensions of the SM, there are many sources of potential LFV \cite{Borzumati:1986qx, Raidal:2008jk, Arganda:2005ji, Gabbiani:1996hi}. In particular, there is the possible presence of off-diagonal soft terms in the slepton mass matrices $(m_L^2)_{ij}$, $(m_{e_R}^2)_{ij}$, as well as in the trilinear couplings $A^e_{ij}$.  In order to construct models that avoid LFV one must introduce strong measures, for example imposing universality \cite{Universality}, but there is often no separate motivation for them.

Most of these approaches rely on a bottom-up approach of building a model which is an effective theory below some scale $\Lambda$. While this approach can be helpful for exploring low-energy phenomenology, it is often hindered by the lack of predictivity associated with the plethora of parameters which are unfixed in such effective models. If, however, one can write a top-down theory, then in principle all the parameters are fixed and calculable up to technical limits. For example, one can consider string/M-theory compactifications which result in the MSSM as the TeV-scale effective theory, with all parameters in principle related and, at least approximately, calculable.

In precisely such an M-theory compactification \cite{Acharya:2007rc, Acharya:2008hi, Acharya:2008zi, Acharya:2011te, Acharya:2012tw} on to a 7D manifold of holonomy group $G_2$, supersymmetry breaking is mediated to the 4D visible sector by gravitational interactions as in minimal supergravity, resulting in a TeV-scale spectrum set by the gravitino mass $M_{3/2}$. A key result of the construction is the derived mini-split between the gauginos, which are suppressed by over an order of magnitude relative to the gravitino mass, and the scalars, which are of order $M_{3/2} \sim \Order(10)$'s of TeV \cite{Acharya:2007rc}. 

In this case, there is minimal flavour violation coming from the superpotential \cite{Acharya:2012tw}. Therefore, any potentially significant flavour-violating effects must either arise due to the flavour structure of the K\"ahler potential or due to renormalisation group running effects from inclusion of the neutrino see-saw mechanism \cite{Hisano:1995cp, Casas:2001sr, Ellis:2001xt}. This will be expounded on further in section \ref{Sources.SEC}.

As with many supergravity-type models, there is uncertainty about the flavour structure of the spectrum due to higher-order corrections to the K\"ahler potential connecting the visible sector to potential hidden sectors in the theory. Because the scalar masses are derived from the K\"ahler potential, their low-energy flavour structure will depend heavily on its flavour structure. If the higher-order corrections are non-flavour-diagonal, they will generally lead to non-universality and non-diagonality of the scalar masses, which in turn will lead to potentially large LFV effects at low energies due to flavour mixing. 

In low-scale supersymmetry models, one is often forced either to require the off-diagonal terms in the K\"ahler potential to be small, or to drive up the scalar masses (see for example \cite{Moroi:2013sfa}). Since the underlying theory we work with here does not allow much variation in the scalar masses, we analyze here whether generic K\"ahler potential structures are allowed, or whether the theory points towards a constrained structure. Indeed we find that in the compactified M-theory framework the constraints from $\mu \to e \gamma$ are already stronger than in a more general theoretical framework. On the other hand, because the high-scale soft breaking Lagrangian is real in M-theory, constraints on EDMs are weaker \cite{Kane:2009kv, Ellis:2014tea}.

In principle the higher-order corrections to the K\"ahler potential should be calculable in a string theory. However, due to limitations in current understanding, such a calculation is difficult. Therefore, we will not attempt such a calculation here. Instead we show how we may use limits from measurements (or non-detection) of low-energy LFV to place constraints on the form of the higher-order corrections and on the connection between the visible and hidden sectors of the theory. In particular, having strong limits on non-diagonality may be a useful probe of potential symmetries and geometry governing flavour dynamics in the visible and hidden sectors.

Since the mechanism for generating small active neutrino masses in the $G_2$-MSSM is currently an open question, we do not consider here the potential LFV effects due to a see-saw type mechanism. If such a mechanism were to exist, depending on the scale of the right-handed neutrinos and the size of the corresponding Yukawa couplings, the LFV effects could be competitive with the LFV coming from the higher-order corrections to the K\"ahler potential. We leave the determination of the precise neutrino mass-generation mechanism and resultant possible LFV effects to future work. Presumably, any LFV effects from neutrino mass mechanisms would generically add to K\"ahler potential effects.

If low-energy LFV were to be measured, then in the limit where all flavour-violation comes from the K\"ahler potential, the form of the higher-order corrections would be known, and could be used to make predictions for other flavour-violating observables. Therefore writing a formalism whereby LFV can be related to the higher-order corrections is of interest. Furthermore, since the higher-order corrections arise due to the connection between the visible and hidden sectors of the theory, determination of the said connection may have important implications for other phenomenological aspects of the theory, such as hidden sector dark matter.

The organisation of this paper is as follows. In section \ref{LFV.SEC} we review the lepton flavour-violating observables which can be used to probe the K\"ahler potential. In section \ref{Sources.SEC}, we review potential sources for LFV in the soft SUSY breaking Lagrangian, and explain why we focus in particular on the K\"ahler potential. In section \ref{Kahler.SEC}, we review the form of the K\"ahler potential and how it may give rise to potential flavour-violating terms in the scalar masses and the trilinear terms. In section \ref{Epsilon.SEC} we analyze the form of the $\sigma$ matrix we introduce in section \ref{Kahler.SEC}, and how its form may be constrained by LFV observables.  Section \ref{Results.SEC} contains the results of numerical computations that show how limits on LFV processes probe the size of the off-diagonal terms in the K\"ahler potential. Finally in section \ref{Conc.SEC} we remark on the significance of the calculation and how future experiments may be able to provide insight on the form of the K\"ahler potential.

\section{Lepton flavour violating processes}
\label{LFV.SEC}
We consider three separate LFV processes, $l_i \rightarrow l_j \gamma$, $l_i\rightarrow 3l_j$ and $\mu-e$ conversion in nuclei. While the process $\mu \rightarrow e \gamma$ is currently the most constrained, significant improvements are expected for the exclusion limits for the other processes also. The current and expected limits are listed in table \ref{Limits.TAB} below.

\begin{table}[H]
\centering
\begin{tabular}{c | c | c}
\textbf{Process} & \textbf{Current limit ($\times 10^{-13}$)} & \textbf{Expected limit ($\times 10^{-13}$)}\\
\hline
$\mu \rightarrow e \gamma$ & $5.7$ \cite{Adam:2013mnn}  & $\sim 6\times10^{-1}$ \cite{Baldini:2013ke}\\
$\mu \rightarrow 3e $ & $10$ \cite{Bellgardt:1987du}  & $\sim 10^{-3}$ \cite{Blondel:2013ia} \\
$\tau \rightarrow \mu \gamma$ & $4.4 \times 10^5$ \cite{Aubert:2009ag} & $\sim 10^4$ \cite{Hayasaka:2013dsa} \\
$\tau \rightarrow e \gamma$ & $3.3 \times 10^5$ \cite{Aubert:2009ag} & $\sim 10^4$ \cite{Hayasaka:2013dsa} \\
$\mu - e$ in Au & $7.0$ \cite{Bertl:2006up} & ? \\
$\mu - e$ in Ti & $17$ \cite{Kaulard:1998rb} & ? \\
$\mu - e$ in Al & ---& $\sim 10^{-3}$ \cite{Comet, Abrams:2012er} \\
\end{tabular}
\caption{Current and future limits on LFV observables.}
\label{Limits.TAB}
\end{table}

\subsection{$l_i \rightarrow l_j \gamma$}
The amplitude of the process $l_i \rightarrow l_j \gamma$ is given by
\beq
T = \epsilon^\alpha \bar{l}_j m_{l_i} i \sigma_{\alpha\beta} q^\beta (A_L P_L +A_R P_R)l_i
\eeq
where $q$ is the momentum and $\epsilon^\alpha$ is the polarisation of the outgoing photon, $m_{l_i}$ is the mass of the incoming lepton $l_i$, $P_{L,R}$ are the usual left and right projection matrices, with $A_{L,R}$ being the coefficients of the amplitude for when the incoming lepton is left/right-handed, and the outgoing lepton is right/left-handed. The dominant one-loop diagrams for this decay are shown in Fig. \ref{1Loop.FIG}, and the expressions for $A_{L,R}$ can be found in Appendix \ref{LFVprocesses.APP}. The branching ratio is then
\beq
\text{BR}(l_i \rightarrow l_j \gamma)=\frac{12\pi^2}{G_F^2} \left( |A_L|^2 + |A_R|^2 \right)
\eeq

\begin{figure}
\centering
\begin{subfigure}[b]{0.4\textwidth}
\includegraphics[scale=0.35]{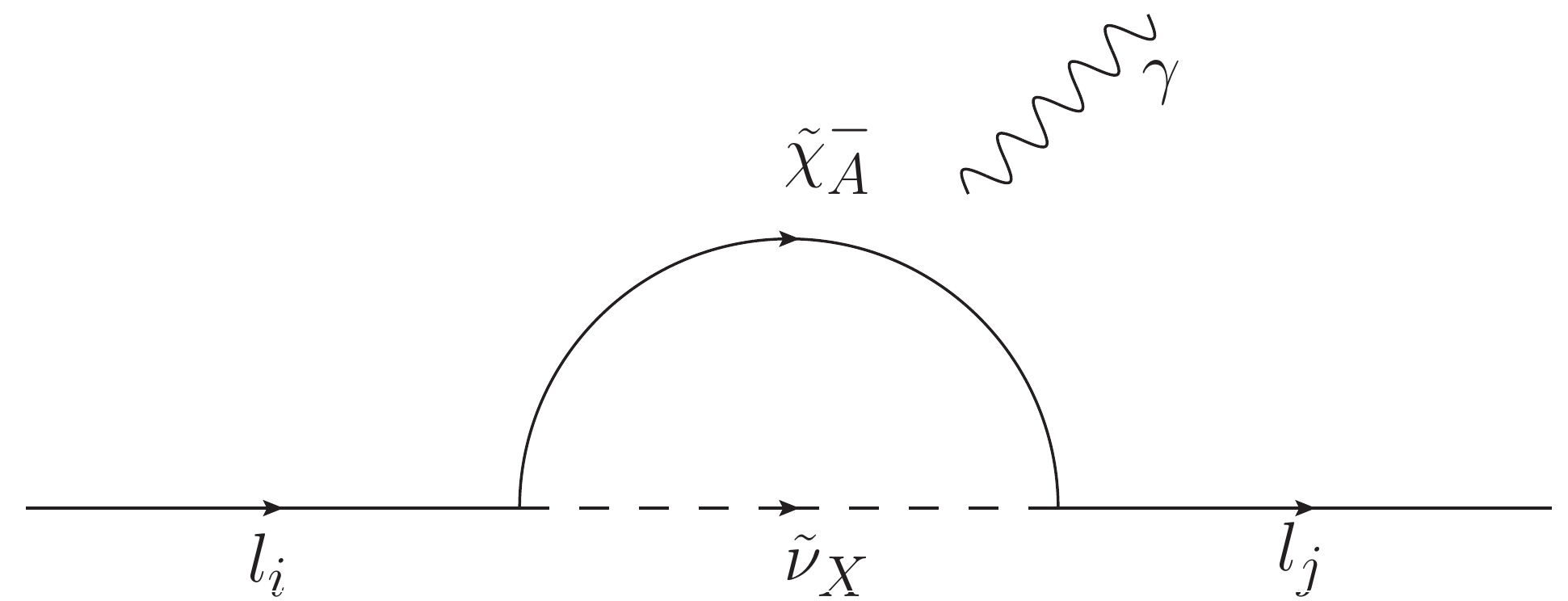}
\end{subfigure}
\hspace{30 pt}
\begin{subfigure}[b]{0.4\textwidth}
\includegraphics[scale=0.35]{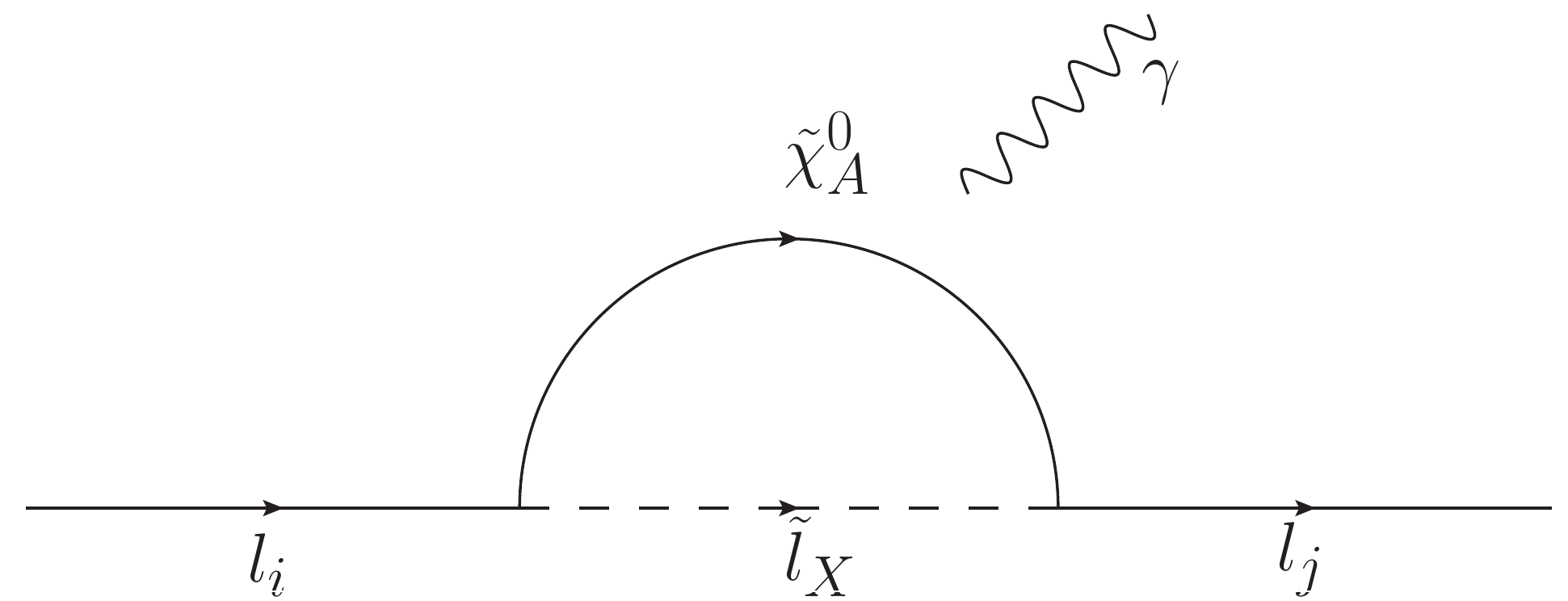}
\end{subfigure}
\caption{Dominant one-loop diagrams contributing to rare $l_i \to l_j \gamma$ decays. $A$ runs from $1,...,4$ for the neutralinos, $1,2$ for charginos, while $X$ runs from $1,...,6$ for the sleptons, and $1,2,3$ for the sneutrinos.}
\label{1Loop.FIG}
\end{figure}

\subsection{$l_i\rightarrow 3l_j$}

We now consider the process $l_i\rightarrow 3l_j$ with particular interest in the process $\mu \rightarrow 3 e$. Contributions to this amplitude come from penguin- and box-type diagrams. The contribution of the penguin-type diagram generally dominates, such that there is a simple relation between $\text{BR}(l_i\rightarrow 3l_j)$ and $\text{BR}(l_i\rightarrow l_j\gamma)$, which is \cite{Hisano:1995cp, Paradisi:2005fk}

\beq
\frac{\text{BR}(l_i\rightarrow 3l_j)}{\text{BR}(l_i\rightarrow l_j\gamma)} \simeq \frac{\alpha}{8\pi} \left(\frac{16}{3} \log  \frac{m_{l_i}}{2m_{l_j}} - \frac{14}{9}\right) \simeq 0.007 \text{ (for $ l_i \equiv \mu $, $l_j \equiv e  $)}
\eeq
Since this process is suppressed relative to $\mu \to e \gamma$, and the future experimental sensitivity will not be a significant improvement on the future sensitivity from $\mu \to e \gamma$ (c.f. Table \ref{Limits.TAB}), we simply make an estimate of the potential future sensitivity. Once a measurement or new limit on $\mu \to e \gamma$ is made, we will be in a position to use the consequent estimate of the size of LFV effects as a guide for a prediction of the $\mu \to 3e$ branching ratio.

\subsection{$\mu-e$ conversion in nuclei}
The conversion of $\mu$ to $e$ in nuclei comes about as a result of Penguin-type diagrams and box-type diagrams \cite{Hisano:1995cp, Paradisi:2005fk}. In general, this process is suppressed relative to the $\mu \to e \gamma$ process \cite{Hisano:1995cp, Paradisi:2005fk}. Additionally, there are hadronic uncertainties which make the theoretical determination of the process less exact, which gives us a less accurate handle on the impact of the higher-order corrections. However, since the future experimental sensitivity from $\mu \to e$ conversion in Al should be an improvement on the sensitivity from experimental searches for $\mu \to e \gamma$, we analyse here the prospect for improving bounds on higher-order corrections from measuring the conversion rate. 

In TeV-scale supersymmetry, the dipole diagrams with photon exchange dominate. However, with a slightly split scenario such as in the G$_{2}$-MSSM or mini-split supersymmetry one must take into account all diagrams, as the penguins may indeed become important.

The branching ratio for $\mu \to e$ conversion in nuclei is given by \cite{Altmannshofer:2013lfa} 
\begin{align}
\nonumber \text{BR}(\mu \to e)_N &= \bigg\{\bigg|\frac{1}{4}A_L^* D+2(2g_{L,V}^u+g_{L,V}^d)V^{(p)}+2(g_{L,V}^u+2g_{L,V}^d)V^{(n)}\bigg|^2 \\&+ \bigg|\frac{1}{4}A_R^* D+2(2g_{R,V}^u+g_{R,V}^d)V^{(p)}+2(g_{R,V}^u+2g_{R,V}^d)V^{(n)}\bigg|^2 \bigg\}\frac{1}{\omega_{capture}}
\end{align}
where $\omega_{capture}$ is the muon capture rate of the nucleus. The $A_{L(R)}$ are the dipole coefficients and $g^{u,d}_{L(R),V}$ are the penguin- and box-type Wilson coefficients coupling to up or down-type quarks as defined in Appendix \ref{LFVprocesses.APP}. The terms $D$, $V^{(p)}$ and $V^{(n)}$ are overlap integrals calculated in \cite{Kitano:2002mt} whose values are presented in Appendix \ref{LFVprocesses.APP} for convenience.

\section{Overview of potential sources of flavour violation}
\label{Sources.SEC}

In general supersymmetric theories, there are many terms which could potentially give rise to flavour-violating observables. The supersymmetry-breaking squark and slepton mass terms and the trilinear terms coupling Higgs to scalars are the main potential sources of flavour violation. Taking the soft supersymmetry-breaking Lagrangian as written in Eq. (\ref{Soft.EQ})
\begin{align}
\label{Soft.EQ}
\nonumber \mathcal{L}^{soft}_{MSSM} &=  \frac{1}{2}\left(M_1 \tilde{B}\tilde{B} + M_2\tilde{W}\tilde{W} + M_3 \sg\sg + h.c. \right) \\ 
\nonumber&+ \left(\tilde{\bar{u}}\bm{A_u} \tilde{Q} H_u - \tilde{\bar{d}} \bm{A_d} \tilde{Q} H_d - \tilde{\bar{e}} \bm{A_e} \tilde{L} H_d + h.c.\right)\\
\nonumber&-\tilde{Q}^{\dagger}\bm{m_Q^2}\tilde{Q} - \tilde{L}^{\dagger}\bm{m_L^2}\tilde{L}-\tilde{\overline{u}} \bm{m_{\bar{u}}^2}\tilde{\overline{u}}^{\dagger} -\tilde{\overline{d}} \bm{m_{\bar{d}}^2}\tilde{\overline{d}}^{\dagger} -\tilde{\overline{e}} \bm{m_{\bar{e}}^2}\tilde{\overline{e}}^{\dagger} \\
&- m_{H_u}^2 H_u^* H_u - m_{H_d}^2 H_d^* H_d - (\mu H_u H_d +h.c.)
\end{align}
we can see that the trilinears $\bm{A_f}$ and the explicit squark and slepton mass terms have the potential to cause mixing between different flavours. 

In the $G_2$-MSSM, the soft terms arise from the superpotential originally derived in \cite{Acharya:2007rc}
\beq
\label{SP.EQ}
W = A_1(\det(\phi^2))^{a/2} e^{-b_1 f} + A_2 e^{-b_2 f} + Y'_{\alpha\beta\gamma}\Phi^\alpha\Phi^\beta\Phi^\gamma
\eeq
and the K\"ahler potential derived by Acharya and Bobkov in section III of \cite{Acharya:2008hi}, including the higher-order term derived in Eq. (62) of \cite{Acharya:2008hi}.
\beq
\label{KP.EQ}
K = -3\log4\pi^{1/3}V_7 +  \kappa_{\alpha\beta}\frac{\Phi^\alpha {\Phi^\beta}^{\dagger}}{V_7} + \frac{\bar{\phi}\phi}{V_7} + c_{\alpha\beta} \frac{\bar{\phi}\phi}{3V_7} \frac{\Phi^\alpha {\Phi^\beta}^{\dagger}}{V_7 } + \ldots
\eeq

The first two terms of the superpotential are for the moduli superpotential, and the third is the MSSM superpotential, with $\Phi^\alpha$ denoting MSSM superfields. The terms $A_i$ are normalisation constants, $b_i$ are the beta function coefficients of the hidden sector and $f$ are the corresponding gauge kinetic functions given by $f=\sum N_i z_i$ where $z_i$ are the complex moduli fields. In the K\"ahler potential, $V_7$ is the volume of the 7D manifold, $\Phi^\alpha$ are visible sector fields, $\phi$ are hidden sector effective fields,arising from chiral fermion condensates, and $\ldots$ represent higher-dimensional operators which are suppressed by higher powers of $V_7$.

Let us consider then where flavour violation may arise in the superpotential written in equation \ref{SP.EQ}. The moduli superpotential does not contain visible sector fields, and therefore cannot result in flavour-violation. The MSSM superpotential can of course give rise to terms that couple different flavours to each other, leading to potentially large flavour violation. The un-normalised Yukawa couplings $Y'_{\alpha\beta\gamma}$ arise from membrane instantons that connect singularities where the chiral superfields exist. They are given by 
\beq
Y'_{\alpha\beta\gamma} = C_{\alpha\beta\gamma}e^{2\pi i V_{Q^{\alpha\beta\gamma}}}
\eeq
where $C_{\alpha\beta\gamma}$ is a matrix of order 1 constants, and $V_{Q^{\alpha\beta\gamma}}$ is the volume of the cycle $Q^{\alpha\beta\gamma}$ connecting the singularities where the chiral superfields exist. In these compactifications, it is natural to obtain a hierarchical structure of these un-normalised Yukawa couplings, due to the dependence on the volume of the cycles, and due to possible family symmetries \cite{Acharya:2008hi}. The normalised Yukawa couplings depend on the K\"ahler metric, and are \cite{Acharya:2008hi}
\beq
Y_{\alpha\beta\gamma}=\frac{\overline{W}}{|W|}e^{K/2} Y'_{\alpha\beta\gamma}(\tilde{K}_\alpha \tilde{K}_\beta \tilde{K}_\gamma)^{-1/2}
\eeq
where $K_\alpha$ is the diagonalised K\"ahler metric in the canonical basis. The dependence on the K\"ahler potential and metric is described in section \ref{Kahler.SEC}. These in turn give rise to the un-normalised trilinear terms of the soft-breaking Lagrangian, which are explictly \cite{Acharya:2008hi}
\beq
\label{Trilinears.EQ}
A'_{\alpha\beta\gamma} = \frac{\overline{W}}{|W|}e^{K/2}F^m \left[K_m Y'_{\alpha\beta\gamma} + \partial_m Y'_{\alpha\beta\gamma} - \left(\tilde{K}^{\delta\rho}\partial_m \tilde{K}_{\rho\alpha} Y'_{\delta\beta\gamma} + \alpha \leftrightarrow \beta + \alpha \leftrightarrow \gamma \right)\right]
\eeq
where $K_m$ is the derivative of the K\"ahler potential with respect to the $m$-th moduli field $s_i$, $K_m = \frac{\partial K}{\partial s_i}$, as originally defined in \cite{Acharya:2008hi}. In the case where the K\"ahler metric $\tilde{K}_{\alpha\beta} \propto \delta_{\alpha\beta}$, the trilinears may then be factorised into
\beq
A'_{\alpha\beta\gamma} = A_{\alpha\beta\gamma}Y_{\alpha\beta\gamma}
\eeq
where the matrix $A_{\alpha\beta\gamma}$ is found by making the explicit computation from Eq. (\ref{Trilinears.EQ}) above. One may then rotate the trilinear terms to the Super-CKM basis, where the squark (and slepton) fields are rotated by the same matrices which rotate the quark (and lepton) fields in the Standard Model. Since the lepton fields can always be rotated such that they are aligned with the gauge interactions, the only potential for flavour violation from rotating the slepton fields is in the case of misalignment between the $A_{\alpha\beta\gamma}$ matrix and the Yukawa matrix. If the decomposition is different from
\beq
A'_{\alpha\beta\gamma} = \bmat a A_0 & & \\ & b A_0 & \\ & & c A_0\emat Y_{\alpha\beta\gamma}
\eeq
where $a,b,c$ are $\Order(1)$ constants, then there is potential flavour violation arising due to the trilinear terms. 

Turning to the K\"ahler potential written in equation (\ref{KP.EQ}), we see that the higher-order corrections may give rise to non-flavour-diagonal terms if there are no symmetries to prevent them. The first term in the potential does not depend on visible sector fields, and therefore will not give rise to flavour-violation. The term only dependent on the visible sector fields, 
\beq
\kappa_{\alpha\beta}\frac{\Phi^\alpha {\Phi^\beta}^{\dagger}}{V_7}
\eeq
will not result in flavour violation either, as we may rotate the fields $\Phi \to \mathcal{U}\Phi$ such that $\mathcal{U}^\dagger \kappa_{\alpha\beta}\mathcal{U} \propto \delta_{\alpha\beta}$. Turning then to the first term containing both visible and hidden sector fields, 
\beq
c_{\alpha\beta} \frac{\bar{\phi}\phi}{3V_7} \frac{\Phi^\alpha {\Phi^\beta}^{\dagger}}{V_7 }
\eeq
we see that there is now a coupling $c_{\alpha\beta}$ that depends on the moduli fields, which is not necessarily proportional to the leading order coupling $\kappa_{\alpha\beta}$. Therefore, when the field rotation is performed, this term may not be proportional to the identity. The coupling $c_{\alpha\beta}$ defines the connection between the hidden sector and the visible sector of the theory, and in principle should be calculable. However, due to technical limitations, it is not yet known precisely what form this coupling takes. Therefore, for generality we cannot assume that it is proportional to $\kappa_{\alpha\beta}$, and can therefore have off-diagonal terms after the rotation of the MSSM fields $\Phi$ is performed. 

If the coupling $c_{\alpha\beta}$ is to be proportional to the leading order coupling $\kappa_{\alpha\beta}$ in the basis where $\kappa_{\alpha\beta}$ is diagonal, it would suggest that there are symmetries that the visible sector fields must obey such that any term with the combination $\Phi^\alpha {\Phi^\beta}^{\dagger}$, regardless of the hidden sector field dependence, must be flavour-diagonal. The matter fields $\Phi$ are charged under the usual MSSM gauge groups, but are also charged under $U(1)$ factors which arise from the Kaluza-Klein reduction of the field in eleven-dimensional supergravity along the two-sphere which shrinks to zero size at the singularity where the chiral field lives \cite{Acharya:2004qe}. It has been argued in \cite{Bourjaily:2009ci} that in local models in M-theory, when there is a branching of a gauge group $G \to H \times U(1)^1 \times \ldots \times U(1)^k$ where the $U(1)^i$ factors have associated charges $q^i$, there can only be a single field with charge $q^i$. This would imply that the only terms allowed in the K\"ahler potential are of the form $\Phi^\alpha {\Phi^\alpha}^{\dagger}$, i.e. the K\"ahler potential would be flavour-diagonal, and would therefore generate flavour-diagonal soft masses. If these symmetries are unbroken, this would preserve the flavour-diagonality of the K\"ahler potential, including the higher-order corrections, and therefore result in zero LFV effects. If however the $U(1)$s are broken at some scale, this would lead to off-diagonal terms in the K\"ahler potential. Such terms would be suppressed as long as the breaking occurs an order of magnitude or more below the Planck scale. We additionally entertain the possibility that there may be geometrical arguments that apply to the matter fields, which could affect the flavour dynamics.



In this section we have given an overview of the possible origins of flavour violation in the compactified String/M-theory framework. In the following section, we go through the calculation of how flavour violation may enter the relevant soft Lagrangian terms from the K\"ahler potential as a result of assuming that the higher-order coupling $c_{\alpha\beta} \not\propto \kappa_{\alpha\beta}$, so that LFV does indeed occur.

Readers, if they wish to only see constraints from LFV, may turn to section \ref{Results.SEC} and treat $\epsilon$ as a small parameter within the compactified framework, directly comparable to the mass insertion parameter $\delta$ often employed in the literature, arising due to the higher-order corrections to the K\"ahler potential.

\section{The K\"ahler potential, and estimates of the scalar masses and trilinears}
\label{Kahler.SEC}

We use the K\"ahler potential derived in \cite{Acharya:2008hi} that we described in detail in section \ref{Sources.SEC}:

\beq
K = -3\log4\pi^{1/3}V_7 +  \kappa_{\alpha\beta}\frac{\Phi^\alpha {\Phi^\beta}^{\dagger}}{V_7} + \frac{\bar{\phi}\phi}{V_7} + c_{\alpha\beta} \frac{\bar{\phi}\phi}{3V_7} \frac{\Phi^\alpha {\Phi^\beta}^{\dagger}}{V_7 } + \ldots
\eeq
where as before, $V_7$ is the volume of the 7D manifold, $\Phi^\alpha$ are visible sector fields, $\phi$ are hidden sector fields and $\ldots$ denotes the higher dimensional operators which are suppressed by higher powers of $V_7$. The term that depends on both the hidden and visible sector fields is a higher-order correction to the K\"ahler potential which couples the two sectors gravitationally. 
 
 \subsection{Scalar masses}
 \label{Scalars.SSEC}
 
 The scalar masses in gravity-mediated models are given by the following equation \cite{Brignole:1997dp}
\beq
\label{scalars.EQ}
m_{\alpha\beta}^2 = (m_{3/2}^2 ) \tilde{K}_{\alpha\beta} - e^{\hat{K}}\bar{F}^{\bar{m}}(\partial_{\bar{m}} \partial_n \tilde{K}_{\alpha\beta} - \partial_{\bar{m}} \tilde{K}_{\alpha\gamma}\tilde{K}^{\gamma\delta}\partial_n \tilde{K}_{\delta\beta})F^n
\eeq
in the unnormalised basis. Here $\tilde{K}_{\alpha\beta} = \partial_\alpha \partial_\beta K$ is the K\"ahler metric of the visible-sector K\"ahler potential $K$ and $F^m$ are the moduli F-terms. One must then canonically normalise the visible matter K\"ahler potential by introducing a normalisation matrix $\mathcal{U}$:
\begin{align}
\Phi \rightarrow ~&\mathcal{U} \Phi
\\
\mathcal{U}^\dagger \tilde{K} \mathcal{U} &= 1
\end{align}
 
 From equation (\ref{scalars.EQ}), we can see that the expression for the scalar masses can then be written as:
\beq
m_{\alpha\beta}^2 = (m_{3/2}^2) \delta_{\alpha\beta} - \mathcal{U}^\dagger \Gamma_{\alpha\beta} \mathcal{U}
\eeq
where 
\beq
\Gamma_{\alpha\beta} = e^{\hat{K}}\bar{F}^{\bar{m}}(\partial_{\bar{m}} \partial_n \tilde{K}_{\alpha\beta} - \partial_{\bar{m}} \tilde{K}_{\alpha\gamma}\tilde{K}^{\gamma\delta}\partial_n \tilde{K}_{\delta\beta})F^n
\eeq

The explicit calculation of the term $\Gamma_{\alpha\beta}$ is in Appendix \ref{ScalarsCalc.APP}. We present the leading-order result for the scalar masses in equation (\ref{ScalarMasses.EQ})
\begin{align}
\label{ScalarMasses.EQ}
\nonumber m_{\alpha\beta}^2 &= M_{3/2}^2 \tilde{K}_{\alpha\beta} - \Gamma_{\alpha\beta}\\
&\approx M_{3/2}^2 \tilde{K}_{\alpha\beta} - \left(m_{1/2}^{tree}\right)^2 \bigg[ \frac{245}{9} - \frac{14P_{eff}}{3} + P_{eff}^2 \bigg] \frac{c\kappa_{\alpha\gamma}\sigma^\gamma_\beta \bar{\phi}\phi}{3V_7^2}
\end{align}
where in the second line we have neglected the terms higher-order in $\frac{\bar{\phi}\phi}{V_7}$ as they are subleading. The parameter $P_{eff}$ is defined in Appendix \ref{ScalarsCalc.APP} and is a constant. This can be reformulated entirely in terms of the gravitino mass as
\beq
m_{\alpha\beta}^2 \approx M_{3/2}^2 \tilde{K}_{\alpha\beta} - \left(-\frac{1}{P_{eff}}\left( 1 + \frac{2V_7}{(Q-P)\phi_0^2}\right)M_{3/2}\right)^2 \bigg[ \frac{245}{9} - \frac{14P_{eff}}{3} + P_{eff}^2 \bigg] \frac{c\kappa_{\alpha\gamma}\sigma^\gamma_\beta \bar{\phi}\phi}{3V_7^2}
\eeq

We made use of the parameterisation of the coupling of the first term involving both hidden and visible sector fields as being
\beq
c_{\alpha\beta} = c \cdot \kappa_{\alpha\delta}\sigma^\delta_\beta
\eeq
where $\sigma^\delta_\beta$ is some unknown matrix that is expected not to depend on the moduli fields and c is a factor defined in \cite{Acharya:2008hi} which can take on a value between 0 and 1. This parameterisation is based on the assumption that the visible sector fields should be coupled in the same way as the leading-order term, by $\kappa_{\alpha\beta}$, with corrections due to the connection to the hidden sector, denoted by $\sigma^{\alpha\beta}$. Therefore, in the limit where symmetries protect the higher-order terms from having flavour off-diagonal terms, this would be equivalent to the statement that $\sigma^{\alpha\beta} \propto \delta^{\alpha\beta}$.

Using the definition of the K\"ahler metric in equation (\ref{Metric.EQ}), we remark that the term proportional to $\cab$ is an order of magnitude suppressed due to the extra factor of $\frac{\bar{\phi}\phi}{V_7}$, relative to the term proportional to $\kab$. Therefore
\beq
\tilde{K}_{\alpha\beta} \simeq \frac{\kab}{V_7}
\eeq
after which we may take the expression for the scalar masses (Eq. (\ref{ScalarMasses.EQ})), and diagonalise to the canonical basis, such that now we have that
\begin{align}
m_{\alpha\beta}^2 &\approx m_{3/2}^2 \delta_{\alpha\beta} - \left(m_{1/2}^{tree}\right)^2 \bigg[ \frac{245}{9} - \frac{14P_{eff}}{3} + P_{eff}^2 \bigg] \frac{c \sigma_{\alpha\beta} \bar{\phi}\phi}{3V_7}
\label{M32dep.EQ}
\end{align}
\begin{align}
\approx m_{3/2}^2 \delta_{\alpha\beta} - \mathcal{C} \sigma_{\alpha\beta}
\end{align}
where $\mathcal{C}$ is a parameter that depends on the value of $c$ only, as all other values are fixed. The size of this parameter can be calculated, using the results of \cite{Acharya:2008hi} that $m_{1/2}^{tree} \sim -0.03 ~\eta~ m_{3/2}$ and $\eta \sim 0.96$, we have that $m_{1/2}^{tree} \sim -0.03 m_{3/2}$. Putting this together with the results of \cite{Acharya:2008hi} that $\frac{\phi_0^2}{V_7} \sim 0.75$, and $P_{eff} \sim 63$, we find
\beq
\mathcal{C} \simeq 0.9 c~ m_{3/2}^2
\eeq
Since the expression for the tree-level gaugino masses is \cite{Acharya:2008hi}
\beq
m_{1/2}^{tree} \approx -\frac{1}{P_{eff}}\left( 1 + \frac{2V_7}{(Q-P)\phi_0^2}\right)M_{3/2}
\eeq
we see that the size of $\mathcal{C}$ is consistent with expectations, and can be up to of order 1 compared with the gravitino mass, since the dependence of $P_{eff}$ cancels in the second term of Eq. (\ref{M32dep.EQ}).
\subsection{Trilinear terms}
\label{Trilinears.SSEC}
The trilinears may also give rise to potential flavour observables. The expression for them is given in equation (190) of \cite{Acharya:2008hi} and is

\beq
A'_{\alpha\beta\gamma} = \frac{\overline{W}}{|W|}e^{K/2}F^m \left[K_m Y'_{\alpha\beta\gamma} + \partial_m Y'_{\alpha\beta\gamma} - \left(\tilde{K}^{\delta\rho}\partial_m \tilde{K}_{\rho\alpha} Y'_{\delta\beta\gamma} + \alpha \leftrightarrow \beta + \alpha \leftrightarrow \gamma \right)\right]
\eeq
in the unnormalised basis. Again, $\alpha,\beta, \gamma$ label visible sector fields and $Y'_{\alpha\beta\gamma}$ are the unnormalised Yukawas in the superpotential. These are defined by the following equation
\beq
Y'_{\alpha\beta\gamma} = C_{\alpha\beta\gamma}e^{i2\pi \sum_i m_i^{\alpha\beta\gamma}z_i}
\eeq
where the coefficients $C_{\alpha\beta\gamma}$ are constants. The integer combination of the moduli $V_{Q^{\alpha\beta\gamma}}= \sum_i m_i^{\alpha\beta\gamma}s_i$ is the volume of the cycle $Q^{\alpha\beta\gamma}$ connecting the singularities $\alpha,\beta,\gamma$ where the chiral multiplets are localised. The physical Yukawa couplings are given by 
\beq
Y_{\alpha\beta\gamma} = \frac{\overline{W}}{|W|}e^{K/2} Y'_{\alpha\beta\gamma}(\tilde{K}_\alpha \tilde{K}_\beta \tilde{K}_\gamma)^{-1/2}
\eeq

The full calculation of the trilinears is found in Appendix \ref{TrilinearsCalc.APP}. We present here the leading order result
\begin{align}
\nonumber A'_{\alpha\beta\gamma} &\simeq \Bigg\{ \left[ \frac{\phi_0^2}{3V_7}\left(Y_{\alpha\beta\gamma} - c\left(1-\frac{7}{3P_{eff}}\right)\left[\frac{\sigma^\delta_\alpha}{3} Y_{\delta\beta\gamma} + \alpha\leftrightarrow\beta + \alpha\leftrightarrow\gamma \right]\right)\right] \\&+ \frac{4\pi V_{Q^{\alpha\beta\gamma}} + 3\lambda}{P_{eff}}Y_{\alpha\beta\gamma}\Bigg\}\left(1+ \frac{2V_7}{(Q-P)\phi_0^2}\right) m_{3/2}
\end{align}

Now we are in a position to comment on the size of potential flavour violation due to the trilinears. The matrix parameterising the misalignment between the leading and higher-order terms in the K\"ahler potential appears with terms of the form $c\sigma^\delta_\alpha Y_{\delta\beta\gamma}/3$. In general, the form of the physical un-diagonalised Yukawa couplings $Y_{\alpha\beta\gamma}$ is not well known, and therefore the potential size of the effect is hard to gauge. However, it is expected that only the couplings of the third generation are of order 1, with all others being smaller \cite{Ramond:1993kv} (also see for example \cite{Ellis:2014tea} and references within).

\section{Form of the $\sigma$ matrix} 
\label{Epsilon.SEC}
Stringy flavour-violating effects will arise from the off-diagonal terms in the $\sigma_{\alpha\beta}$ matrix. Having derived the form of the scalar masses and the trilinears in the case where the higher-order corrections are misaligned relative to the leading order by some matrix $\sigma_{\alpha\beta}$, we now wish to put some limits on the form and size of said matrix. In principle it is a $17 \times 17$ matrix as it runs over $\alpha,\beta = Q_i,~ U_i,~ D_i,~ L_i,~ E_i,~ H_u,~ H_d$ where $i=1,2,3$, but certain restrictions simplify this substantially. Since the elements will not depend on fields that are charged under any of the Standard Model gauge groups, we can restrict ourselves to five $3\times3$ sub-matrices $\sigma^F_{ij}$ where $F=Q,~U,~D,~L,~E$ and $i,j = 1,2,3$. We rule out mixing between the Higgs fields and the $F$ fields, for example by assuming R-parity conservation. Then we are dealing with a matrix of the form
\beq
\sigma_{\alpha\beta} = 
\bmat
\bmat\sigma^Q_{ij}\emat_{3\times3} &  &  &  & &  &  \\
 & \bmat\sigma^U_{ij}\emat_{3\times3} &  & & & & \\
&  & \bmat\sigma^D_{ij}\emat_{3\times3} &  &&&\\
&& &\bmat\sigma^L_{ij}\emat_{3\times3}& && \\
&&&& \bmat\sigma^E_{ij}\emat_{3\times3} &  & \\
 &&& &&\sigma^{H_u}&\\
 &&&&&&\sigma^{H_d}
\emat
\eeq
where $\sigma^{H_u}$ and $\sigma^{H_d}$ are just numbers.

We are now in a position to write down what the scalar mass matrices are at the high scale. We will have four $6\times6$ matrices mixing the left- and right-handed sfermions:
\beq
\bmat \tilde{f}^\dagger_{i,L} & \tilde{f}^\dagger_{i,R} \emat \bmat (m_{3/2}^2 \delta_{ij} - \mathcal{C} \sigma^{F_L}_{ij}) & v(A^*_{ij} \sin\beta - \mu Y_{ij} \cos\beta) \\ v(A_{ij} \sin\beta - \mu^* Y_{ij} \cos\beta) & (m_{3/2}^2 \delta_{ij} - \mathcal{C} \sigma^{F_R}_{ij})\emat \bmat \tilde{f}_{j,L} \\ \tilde{f}_{j,R}\emat
\eeq
where $F_L = Q,~L$, $F_R = U,~D,~E$, and $f_i = u_i,~d_i, \ell_i, \nu_i$. We see that when  $\sigma^F_{ij} \not\propto \delta_{ij}$, this will result in flavour-violating processes. 

Considering that the off-diagonal terms must to be small relative to the diagonals to comply with experimental data, we may treat the $\sigma_{\alpha\beta}$ matrix as being approximately diagonal up to small perturbations. This allows us to employ the results of the calculation of the Higgs mass \cite{Kane:2011kj, Kane:2012qr}, as well as requiring consistent electroweak symmetry breaking (EWSB), to constrain the product of $0.9\times c\sigma_{H_u,H_d}$. 

The Higgs mass is given by
\beq
M_h^2 = M_Z^2 \cos^2 2\beta + \delta M_h^2
\eeq
where $\delta M_h^2$ denotes the radiative corrections from SM and MSSM particles. The value of $\delta M_h^2$ depends on the given values of $M_{3/2},~\mu$ and $0.9 c \sigma_{H_u,H_d}$. Meanwhile the EWSB conditions are
\begin{align}
&\frac{M_Z^2}{2} = \frac{\bar{m}_{H_d}^2 - \bar{m}_{H_u}^2 \tan^2 \beta}{\tan^2 \beta - 1} - \mu^2
\label{EWSB.EQ}
\end{align}
\begin{align}
&B\mu = \frac{1}{2} \sin 2 \beta(\bar{m}_{H_u}^2 + \bar{m}_{H_d}^2 + 2\mu^2)
\end{align}
where bars denote that the soft masses are tadpole corrected, and the evaluation of all parameters is one at the scale $Q_{EWSB} = \sqrt{m_{\tilde{t}_1} m_{\tilde{t}_2}}$.
In our class of models, $\mu < 0.1 M_{3/2}$ is expected \cite{Acharya:2011te}, and $0 < \bar{m}_{H_u}^2 \ll \bar{m}_{H_d}^2$, as $m_{H_u}^2$ runs significantly more than $m_{H_d}^2$ due to the top Yukawa coupling for moderate $\tan \beta \lesssim 10$. Since $m_{H_{u,d}}^2 = M^2_{3/2}(1-0.9 c \sigma_{H_u,H_d})$, by taking appropriate limits one can obtain the following expression for $\mu$ \cite{Ellis:2014kla}
\beq
\mu^2 \approx \frac{M_{3/2}^2(1-0.9 c \sigma_{H_d})}{B^2 - M_{3/2}^2(1-0.9 c \sigma_{H_d})}\left( M_{3/2}^2 f(0.9 c \sigma_{H_u}) + \frac{M_Z^2}{2}\right)
\eeq
where $f(0.9 c \sigma_{H_u})$ takes into account the running of $m_{H_u}^2$.
This equation reduces the three-dimensional $M_{3/2},~\mu,~0.9 c \sigma_{H_u,H_d}$ parameter space to a two-dimensional one. When combined with requiring the experimentally determined Higgs mass, this is further reduced to a one-dimensional strip in parameter space. 

The result assuming a diagonal higher-order correction matrix found in \cite{Ellis:2014kla} is that $0.9\times c\sigma_{H_u,H_d} \simeq 0.5$ for the preferred range of values of $M_{3/2}$. Since the $\sigma_{\alpha\beta}$ matrix depends only on the moduli and hidden sector fields, it should not a priori know about differences between the chiral superfields. Therefore by constraining two elements in the diagonal, we make the well motivated extrapolation that we are in fact constraining the diagonal elements of the full matrix. We therefore make the redefinition that the diagonal of the $\sigma_{\alpha\beta}$ matrix is unity, such that $0.9c \simeq 0.5$. We may then write
\beq
\label{EpsilonMat.EQ}
\sigma_{\alpha\beta} = \delta_{\alpha\beta} + \bmat
\bmat\varepsilon^Q_{ij}\emat_{3\times3} &  &  &  & &  &  \\
 & \bmat\varepsilon^U_{ij}\emat_{3\times3} &  & & & & \\
&  & \bmat\varepsilon^D_{ij}\emat_{3\times3} &  &&&\\
&& &\bmat\varepsilon^L_{ij}\emat_{3\times3}& && \\
&&&& \bmat\varepsilon^E_{ij}\emat_{3\times3} &  & \\
 &&& &&0&\\
 &&&&&&0
\emat
\eeq
with $i\neq j,~i,j=1,2,3$, and $\sigma^F_{ij} = \delta_{ij}+\varepsilon^F_{ij}$. Equation (\ref{EpsilonMat.EQ}) is the key result of the arguments from the compactification and the symmetries involved.

This allows us to study the various limits on the entries of the $\varepsilon^F_{ij}$ matrices, whether in the quark or leptonic sectors, as well as the interplay between the two sectors. Since we are interested in very rare processes, and any uncertainties can be problematic, we focus on leptonic processes, which are subject to fewer uncertainties. Additionally, decays such as $b \to s \gamma$ are relatively large compared to $\mu \to e \gamma$, and the experimental uncertainties on the measurement are many orders of magnitude larger than the potential effect due to the higher-order corrections from the K\"ahler potential. The most important constraint from the quark sector comes from the neutral kaon mass difference $\Delta m_K$, which we compare with the constraints from LFV effects. We verified numerically that the recent experimental measurement of the rare decays of B mesons into muon pairs \cite{CMS:2014xfa} is consistent with our framework. Any strengthening of the constraints from leptonic processes could be applied to the quark sector. In particular, any determination of non-zero LFV effects could be used in the future to make predictions of quark-flavour-violating effects.

We consider a few particular textures for the $3\times3$ matrices $\varepsilon^{L,E}_{ij}$, which we take to be real, characterized by a small parameter $\epsilon$. This parameter serves as an indicator of the scale at which potential flavour symmetries preventing off-diagonal K\"ahler potential couplings are broken, or of geometrical suppression of the off-diagonal couplings.
\begin{align}
\varepsilon^{L,E}_{ij} = \bmat 0&\epsilon & \epsilon \\ \epsilon & 0 & \epsilon \\ \epsilon & \epsilon & 0 \emat,~
\varepsilon^{L,E}_{ij} = \bmat 0&\epsilon & 0 \\ \epsilon & 0 & 0 \\ 0 & 0 & 0 \emat, ~
\varepsilon^{L,E}_{ij} = \bmat 0&0 & \epsilon \\ 0 & 0 & 0 \\ \epsilon & 0 & 0 \emat, ~
\varepsilon^{L,E}_{ij} = \bmat 0&0 & 0 \\ 0 & 0 & \epsilon \\ 0& \epsilon  & 0 \emat
\end{align}

 Texture 1 corresponds to all symmetries being broken at the same scale. Texture 2 corresponds to only the symmetries preventing $\mu e$ mixing being broken, while textures 3 and 4 correspond to the symmetries preventing $\tau e$ and $\tau \mu$ mixing being broken. While texture 1 could in principle have different parameters corresponding to $\mu e$, $\tau e$, and $\tau \mu$ mixing, this would correspond to breaking the various symmetries at different scales, which would then have to be explained at a fundamental level. Therefore, we assume that all the symmetries are broken at a unique scale for texture 1. We study the rare decay of $\mu \to e \gamma$ and $\mu \to e$ conversion in Aluminium in the case of the first two textures, $\tau \to e \gamma$ for the third, and $\tau \to \mu \gamma$ for the fourth. We do not explicitly study the rare tau decays for the first two textures, as the limits due to the $\mu \to e \gamma$ exclusion are already stronger than any limit one could impose from the measurement of tau decays.

\section{Numerical results}
\label{Results.SEC}
In this section we present the numerical results for the various textures and resultant LFV observables we consider. The construction is such that the running of the first two generation scalars is not substantial. This was verified both semi-analytically using the RGEs as found in \cite{Martin:1993zk} and numerically using the spectrum calculator \verb SoftSusy  \cite{Allanach:2001kg}. Therefore, we take all values as being at the electroweak scale in the following calculations.

In many cases in the literature, the branching ratios are calculated in the mass insertion approximation. In the most straightforward application of this approximation, the branching ratio for $l_i \to l_j \gamma$ is given by \cite{Paradisi:2005fk}
\beq
BR(l_i \to l_j \gamma)_{MI} \simeq \frac{\alpha^3}{G_F^2}\frac{\delta^2}{M_S^4}\tan^2\beta
\eeq
where we have defined as an approximately typical supersymmetry scale $M_S = m_0$, where $m_0 = \sqrt{M_{3/2}^2(1-0.9~c)}$ is the universal scalar mass, with $0.9c=0.5$ as found in \cite{Ellis:2014kla}, and the factor of 0.9 was derived in Section \ref{Kahler.SEC}. Here the mass insertion parameter $\delta$ is defined as
\beq
\delta = \left(\frac{\Delta m_{\tilde{l}}^2}{m_{\tilde{l}}^2}\right) = \frac{M_{3/2}^2 0.9c \epsilon}{M_{3/2}^2(1-0.9c)} \sim \epsilon
\eeq
for $0.9c=0.5$.
The $\tan^2\beta$ enhancement comes from the Higgsino contribution to the diagrams in Fig. \ref{1Loop.FIG} \cite{Hisano:1998cx}. While this simple approximation does not give correct results due to the mini-split between the scalars and the gauginos, there exist in the literature more complex expressions in the mass insertion approximation that take the split spectra possibility into account, such as \cite{Altmannshofer:2013lfa}, or more recently \cite{Calibbi:2015kja}. Additionally, while the simple approximation is not useful for calculations, it provides a useful guide as to the parametric dependence of the branching ratio.

As briefly described above, an advantage of the compactified M-theory is that $\mu$ and $\tan\beta$ are related by the theory rather than being free parameters. $\mu$ is incorporated following Witten's approach \cite{Witten:2001bf} and using a generic discrete symmetry. After moduli stabilisation it is estimated \cite{Acharya:2011te} that $\mu \approx \langle moduli\rangle M_{3/2}$, where $\langle moduli \rangle$ is a typical moduil vev, of order $0.1$. Constraining $\mu$ to not arise from the superpotential, and using the EWSB conditions gives (at the high scale) $2 \mu \tan \beta \approx M_{3/2}$. Combining these gives $\tan\beta \simeq 5-10$.

\begin{figure}[H]
\centering
\includegraphics[scale=0.28]{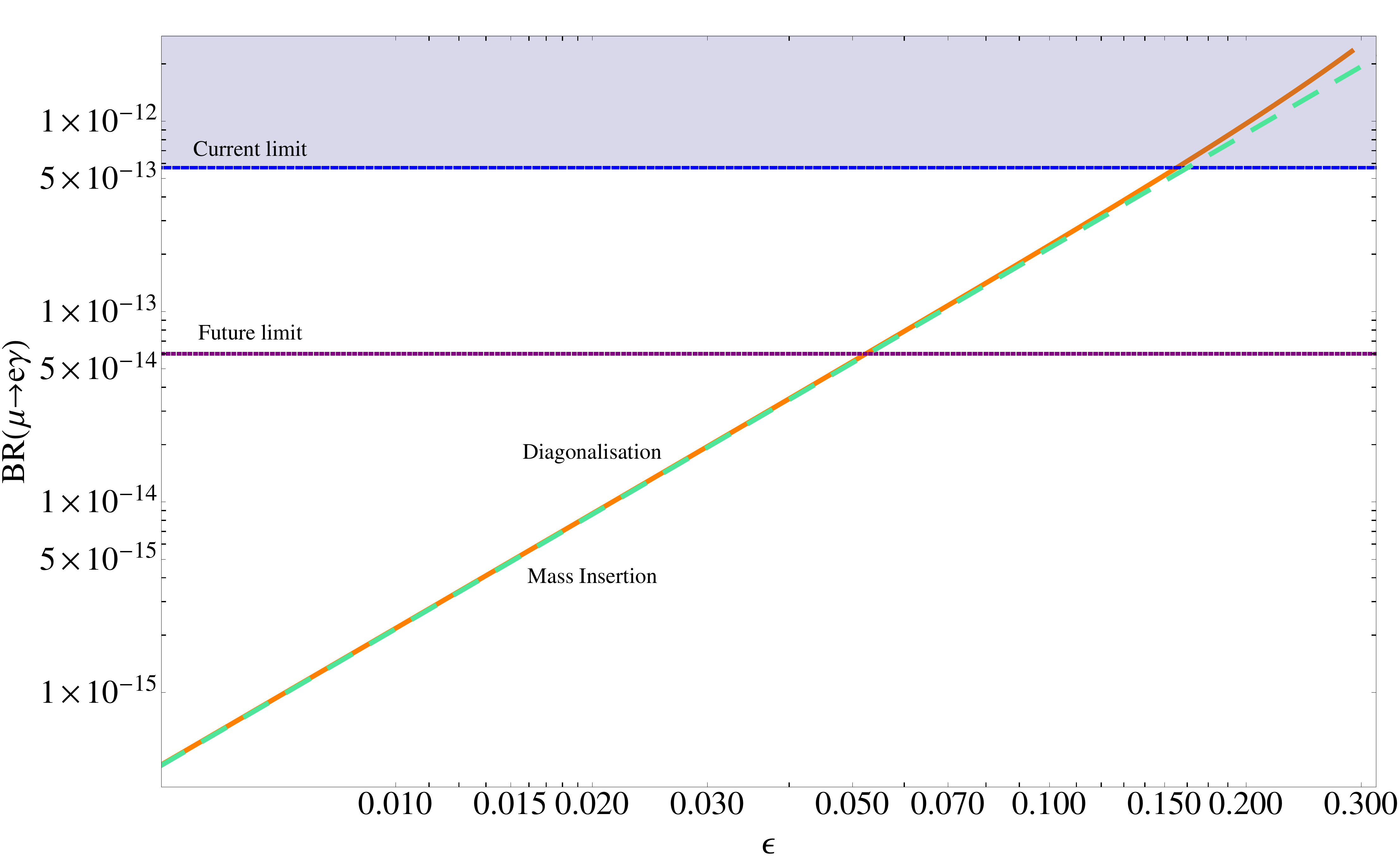}
\caption{Branching ratio of $\mu \to e \gamma$ as a function of the small parameter $\epsilon$ for texture 2. The curves represent the eigenstate basis calculation (orange) and mass insertion approximation (dashed blue) branching ratios for $M_{3/2} = 35$ TeV ($m_0\simeq 24$ TeV), $M_{1} = 400,~M_{2}=600$ GeV, $\mu = 1.4$ TeV, and $\tan\beta=7$. The shaded blue region is excluded by the current experimental limit, shown as a dotted blue line, and the dotted purple line is the expected future sensitivity.}
\label{ComparisonMI.FIG}
\end{figure}

Comparing the result from \cite{Altmannshofer:2013lfa} in the mass insertion approximation with the full diagonalisation result, we obtain Fig. \ref{ComparisonMI.FIG}. 
We note that for larger values of $\epsilon$ there is some discrepancy between the mass insertion approximation and the eigenstate calculation, of order a few percent.

In the subsequent calculations of BR($\mu \to e \gamma$) we perform the calculation in the eigenstate basis, since this gives us discriminatory power between textures 1 and 2, as the current probes are in the relatively large $\epsilon$ region. However, for $\mu \to e$ conversion we use the results of \cite{Altmannshofer:2013lfa} in the mass insertion approximation, since the region probed is at small $\epsilon$, where the two approaches match well.

\subsection{Relevance of $\Delta m_K$}

Since we do not make any a priori assumptions about the form of the higher order corrections to the K\"ahler potential, other than that they should obey symmetries, we must assume that any higher order corrections affecting lepton flavour-violating processes may also impact quark flavour observables. Thus, in the absence of large phases \cite{Kane:2009kv}, we must pay particular attention to the stringent constraints from the small mass splitting of the neutral K mesons, where we consider the case where the supersymmetric contribution to the mass splitting saturates the bound \cite{Agashe:2014kda}, 
\beq 
\Delta m_K <  3.484\times10^{-12} ~\text{MeV}
\eeq
Therefore, we use this constraint to set a limit on the maximal size of the same $\epsilon$ parameter that enters in the leptonic sector. The contribution from supersymmetry has been computed before (see for example \cite{Universality}, and \cite{Gabbiani:1996hi}). We use the next to leading-order (NLO) calculation of \cite{Bagger:1997gg, Ciuchini:1998ix} to compute the relevant limits for our choices of parameters. While we tabulate here the limits, they are also plotted in the figures in the subsequent sections. Note the current limits on $\epsilon$ from $\mu \to e \gamma$ or $\mu \to e$ conversion (c.f. Table \ref{EpsLimits.TAB}) will be the most stringent if there is no discovery.

\begin{table}
\centering
\begin{tabular}{c c c}
\hline
$M_{3/2}$ (TeV) & $m_0$ (TeV) & $\epsilon$ \\
\hline
25 & 17 & 0.09 \\
35 & 24 & 0.14 \\
50 & 35 & 0.22  \\
\hline
\end{tabular}
\caption{Limits on $\epsilon$ due to $\Delta m_K$ for $m_{\tilde{g}}=1.5$ TeV.}
\label{DmKlimits.TAB}
\end{table}

\subsection{Importance of different parameters}
We show that the most important parameters are the gravitino mass and the small parameter $\epsilon$ that appears in the off-diagonal entries of the $\sigma_{\alpha\beta}$ matrices. While the LFV branching ratios depend also on the electroweakino masses $M_a$ ($a=1,2$) and the Higgs bilinear term $\mu$, as can be seen from figs. \ref{Gaugino.FIG} and \ref{Mu.FIG} below, the dependence is only a change of order a few whereas, as will be seen subsequently, the dependence on the gravitino mass $M_{3/2}$ is considerably stronger. 

\begin{figure}[H]
\centering
\begin{subfigure}[t]{0.48 \textwidth}
\includegraphics[width=\textwidth]{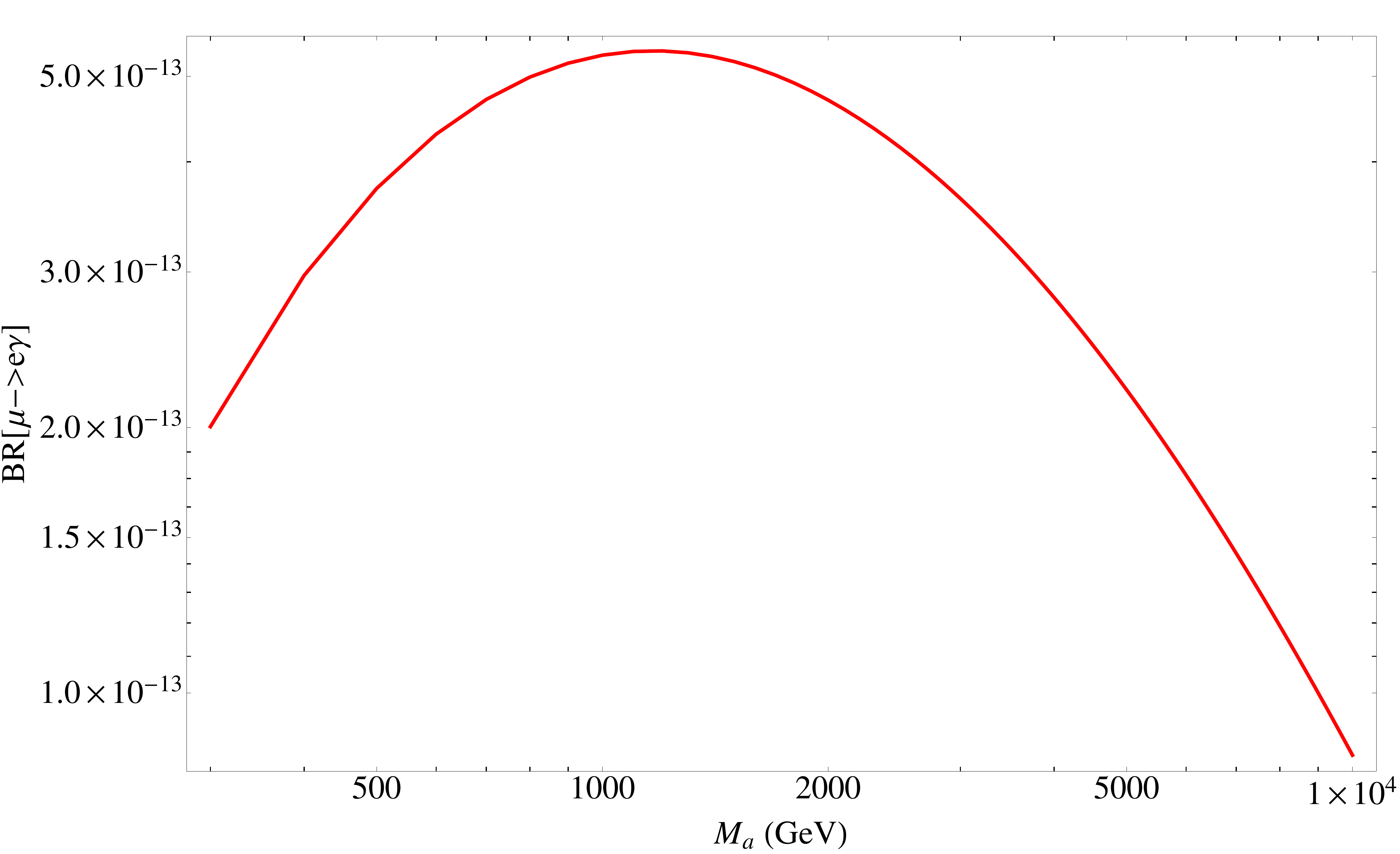}
\caption{The branching ratio of the decay $\mu \to e \gamma$ for texture 1, varying $M_{a}$ between 300 and $10^4$ GeV, using the parameterisation that $M_1=M_a$, $M_2=1.5\times M_a$. The gravitino mass was chosen to be $M_{3/2}=35$ TeV (i.e. $m_0 = 24$ TeV), $\mu = 1.4$ TeV, and $\epsilon =0.1$, to illustrate a set of parameters which would give variations near the current limit, which is just above the range shown here (c.f. Table \ref{Limits.TAB}).}
\label{Gaugino.FIG}
\end{subfigure}
~~
\begin{subfigure}[t]{0.48 \textwidth}
\includegraphics[width=\textwidth]{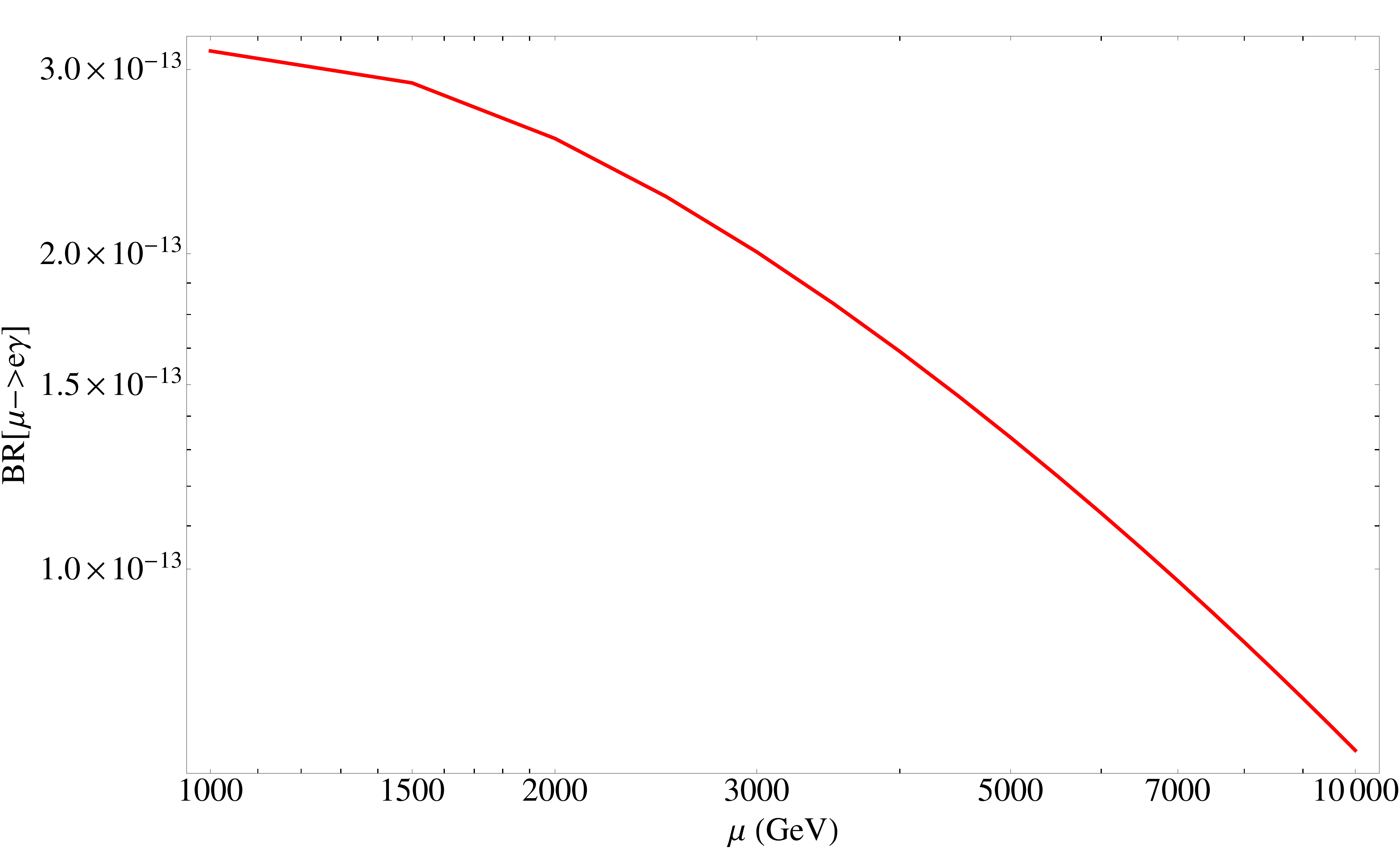}
\caption{The branching ratio of the decay $\mu \to e \gamma$ for texture 1, varying $\mu$ between 300 and $10^4$ GeV. The gravitino mass was chosen to be $M_{3/2}=35$ TeV (i.e. $m_0 = 24$ TeV), and $\epsilon =0.1$, with $M_1 = 400$ and $M_2 = 600$ GeV. This set of values gives values of $BR(\mu \to e \gamma)$ below the current limit (c.f. Table \ref{Limits.TAB}).}
\label{Mu.FIG}
\end{subfigure}
\caption{The dependence of the branching ratio of $\mu \to e \gamma$ on $M_{1,2}$ and $\mu$.}
\end{figure}

In figs. \ref{Gaugino.FIG} and \ref{Mu.FIG} we have considered a larger range of values for the electroweakino masses and $\mu$ than expected from the M-theory compactification to emphasise the lack of sensitivity to these parameters. Thus, variations in the order of magnitude of the electroweakino masses and the value of $\mu$ will alter the result by up to $\Order($few$)$ factors, the overall order of magnitude for the various branching ratios is set by the size of $\epsilon$ and $M_{3/2}$. 

\subsection{Muon decay constraints on textures 1 and 2}

\begin{figure}[H]
\centering
\begin{subfigure}[t]{0.48 \textwidth}
\includegraphics[width=\textwidth]{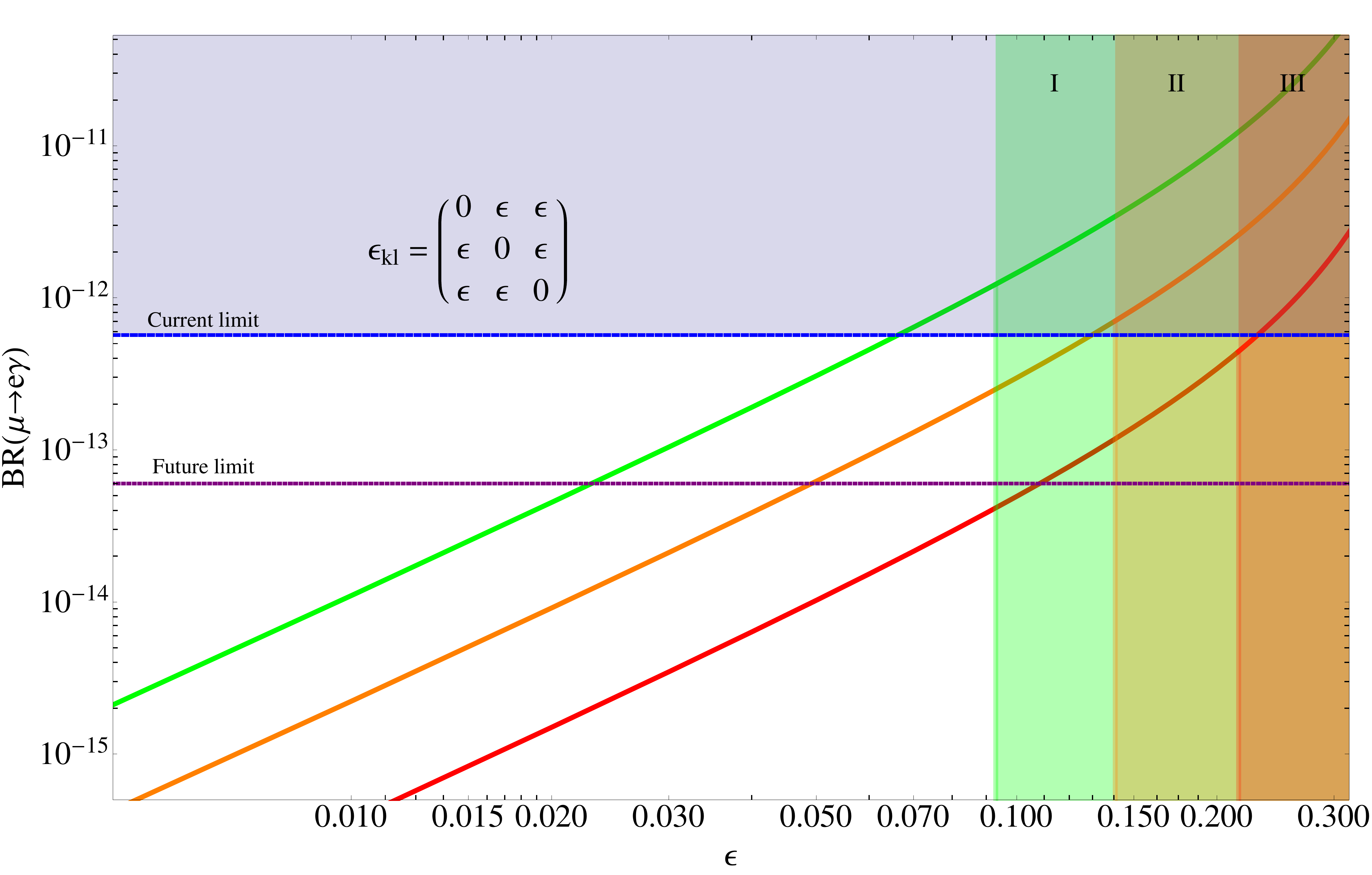}
\caption{Branching ratio of $\mu \to e \gamma$ as a function of the small parameter $\epsilon$ for texture 1. The three curves represent $M_{3/2} = 25,~35,~50$ TeV for green, orange and red respectively. The dotted blue line represents the current experimental limit, and the dotted purple line is the future limit. The shaded regions I, II and III denote the exclusions due to $\Delta{m_K}$ constraints for $M_{3/2} = 25,~35,~50$ TeV respectively.}
\label{fixedM32.FIG}
\end{subfigure}
~~
\begin{subfigure}[t]{0.48 \textwidth}
\includegraphics[width=\textwidth]{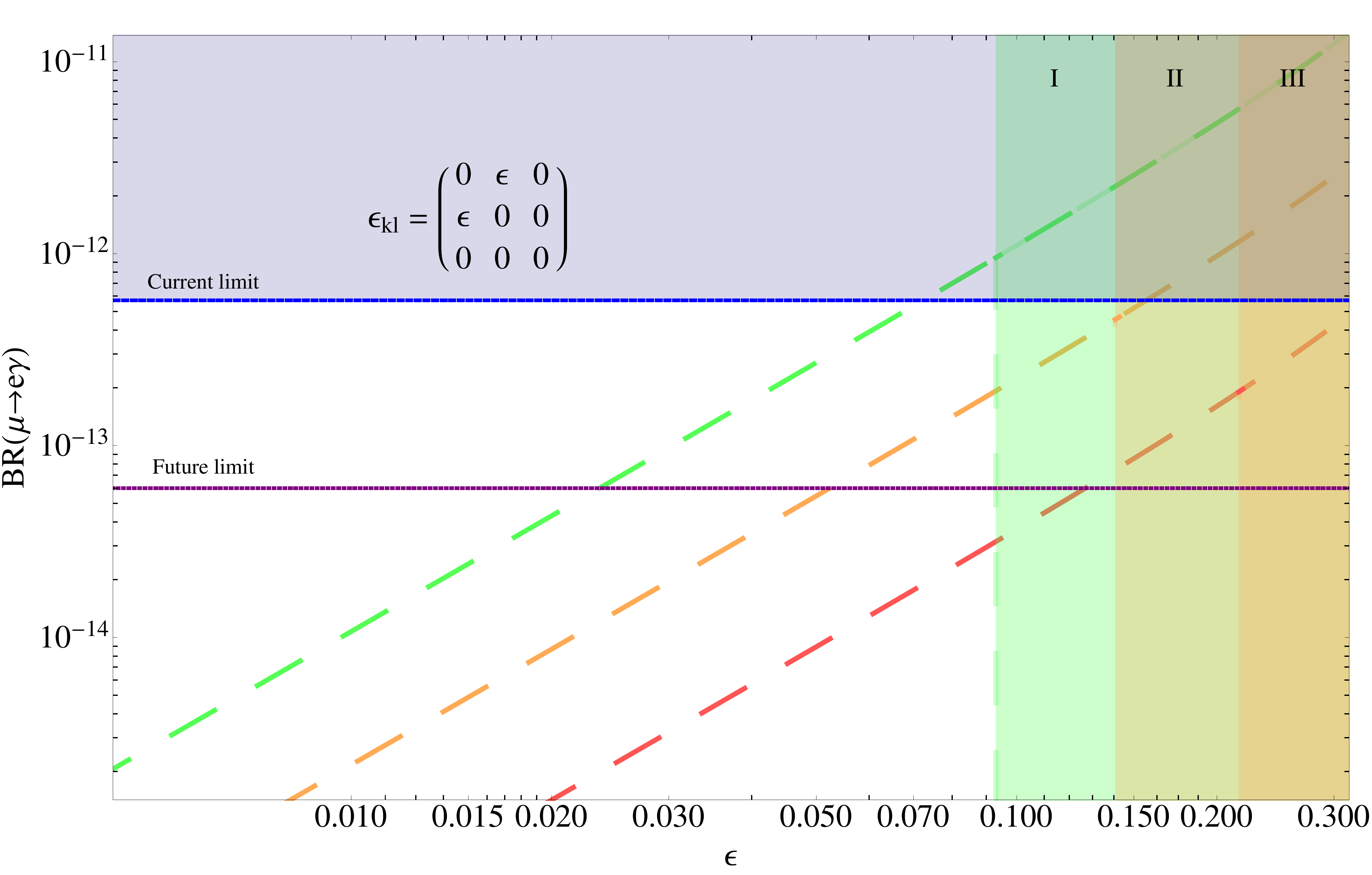}
\caption{Branching ratio of $\mu \to e \gamma$ as a function of the small parameter $\epsilon$ for texture 2. The three curves represent $M_{3/2} = 25,~35,~50$ TeV for green, orange and red respectively. The dotted blue line represents the current experimental limit, and the dotted purple line is the future limit. The shaded regions I, II and III denote the exclusions due to $\Delta{m_K}$ constraints for $M_{3/2} = 25,~35,~50$ TeV respectively.}
\label{fixedM32t2.FIG}
\end{subfigure}
\caption{Limits on $\epsilon$ in the cases of textures 1 and 2 coming from BR($\mu \to e \gamma$). In both plots we fix $\mu = 1.4$ TeV, $m_{\tilde{g}}=1.5$ TeV, $M_1 = 400$ GeV, $M_2 = 600$ GeV and $\tan\beta = 7$.}
\end{figure}

It is clear from the form of texture 1 that this will result in a non-zero $\mu \to e \gamma$ branching ratio, as well as non-zero $\tau \to e \gamma$ and $\tau \to \mu \gamma$ branching ratios. However, since the limits on the tau decays are not as strong as those on the muon decay, we need only consider the muon decay to set limits on the size of $\epsilon$ for various choices of the gravitino mass $M_{3/2}$. This is shown in figure \ref{fixedM32.FIG}.

We can see that for the choices of $M_{3/2} = 25,~ 35$ and $50$ TeV, corresponding to $m_0 = 17,~24$ and $35$ TeV, the current experimental sensitivity on BR$(\mu \to e \gamma)$ results in bounds of $\epsilon \lesssim 0.06,~0.14$ and $0.24$ respectively. The future experimental limit promises to strengthen those bounds to $\epsilon \lesssim 0.02,~0.05$ and $0.11$ if there is no discovery.

\begin{figure}[H]
\includegraphics[scale=0.4]{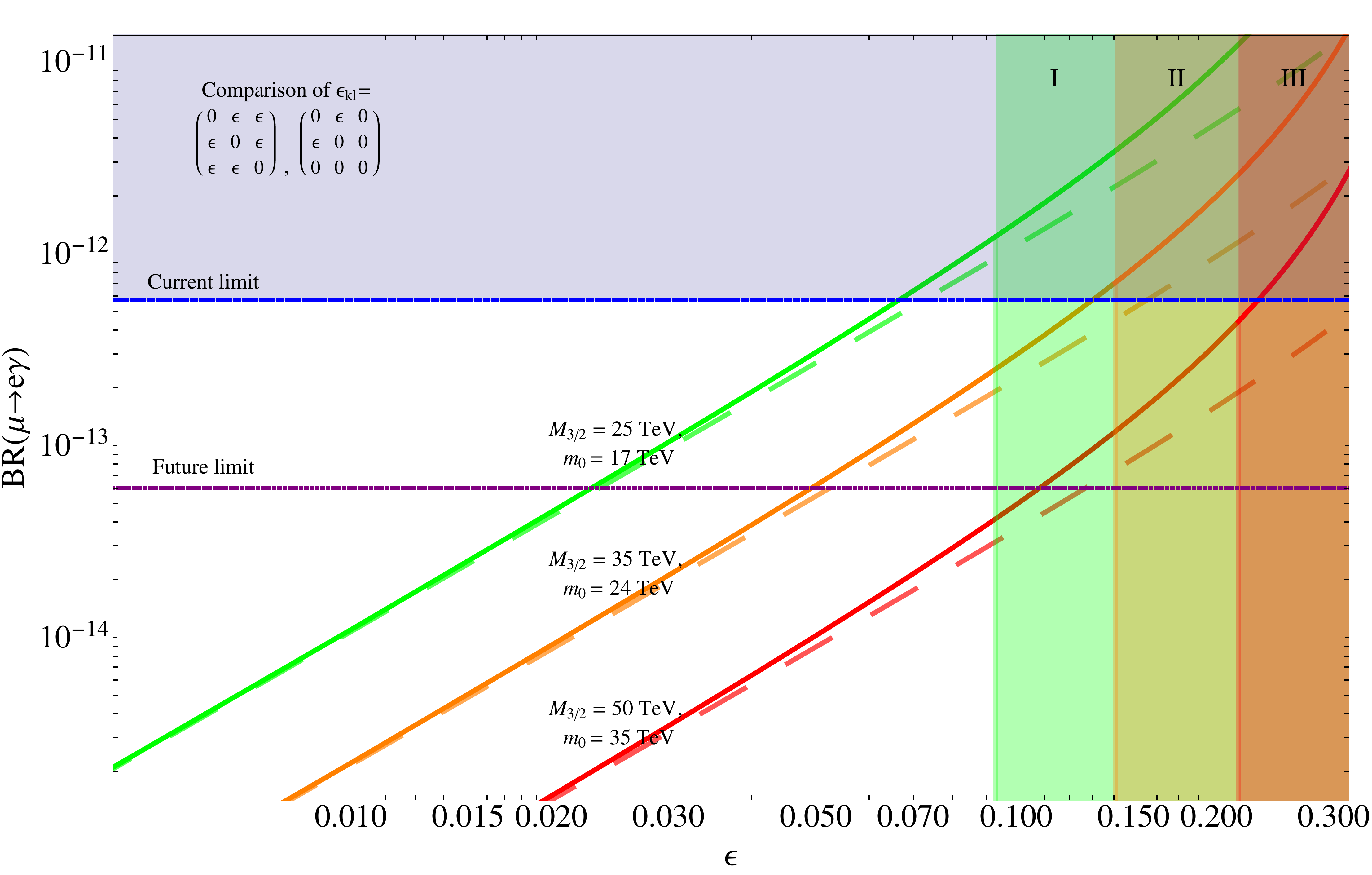}
\caption{Branching ratio of $\mu \to e \gamma$ as a function of the small parameter $\epsilon$ for textures 1 and 2. The curves represent $M_{3/2} = 25,~35,~50$ TeV for green, orange and red respectively. Texture 2 results are shown with dashed lines. The corresponding values of $m_0$ are $17,~24,~35$ TeV. The dotted blue line represents the current experimental limit, and the dotted purple line is the future sensitivity. The blue shaded region is excluded for all values of $M_{3/2}$ by the current experimental limit. The shaded regions I, II and III denote the exclusions due to $\Delta{m_K}$ constraints for $M_{3/2} = 25,~35,~50$ TeV respectively. We fix $\mu = 1.4$ TeV, $m_{\tilde{g}}=1.5$ TeV, $M_1 = 400$ GeV, $M_2 = 600$ GeV and $\tan\beta = 7$.}
\label{Comparison.FIG}
\end{figure}

The results are for texture 2 are shown in figure \ref{fixedM32t2.FIG}. We see that for the choices of $M_{3/2} = 25,~ 35$ and $50$ TeV, corresponding to $m_0 = 17,~24$ and $35$ TeV, the current experimental sensitivity on BR$(\mu \to e \gamma)$ results in bounds of $\epsilon \lesssim 0.08,~0.19$ and $0.5$ respectively. The future experimental sensitivity promises to strengthen those bounds to $\epsilon \lesssim 0.03,~0.06$ and $0.15$ if no discovery is made. The bounds are therefore weaker than those set by texture 1. This can be seen in figure \ref{Comparison.FIG} where we overlay the two textures for the same gravitino mass choices. 

Comparing with the constraints from $\Delta m_K$, we see that the constraints from $\mu \to e \gamma$ are already stronger for $M_{3/2} \lesssim 50$ TeV for texture 1, albeit barely for $M_{3/2}=35$ TeV. For texture 2, we see that BR($\mu \to e \gamma$) is more constraining than $\Delta m_K$ for $M_{3/2} < 35$ TeV only. The reason the sensitivity does not scale quite the same way for the two textures is that in texture 1 there are more avenues for $\mu-e$ flavour violation than in texture 1. Thus, texture 1 exhibits deviation from the naive $\epsilon^2$ dependence for smaller values of $\epsilon$ than texture 2.

If $\epsilon$ were larger than the values quoted above for textures 1 and 2, the decay of $\mu \to e \gamma$ would have been observed, or excluded by $\Delta m_K$. With no currently known mechanism by which to understand why $\epsilon$ should be much smaller than these bounds, this suggests one of three outcomes. The first is that the rare decay of $\mu \to e \gamma$ will be seen in the not too distant future, because $\epsilon$ should not be too much smaller without being contrived. The second is that some as yet not well understood partial symmetry sets a smaller, but non-zero, size for $\epsilon$, whose exploration would require a future generation of experiments. The third is that the K\"ahler potential is, perhaps surprisingly, flavour-diagonal or nearly so, in which case any flavour-violating effects must arise due to some other mechanism.

\subsection{Tau decays as a probe of textures 3 and 4}
\begin{figure}[H]
\includegraphics[scale=0.4]{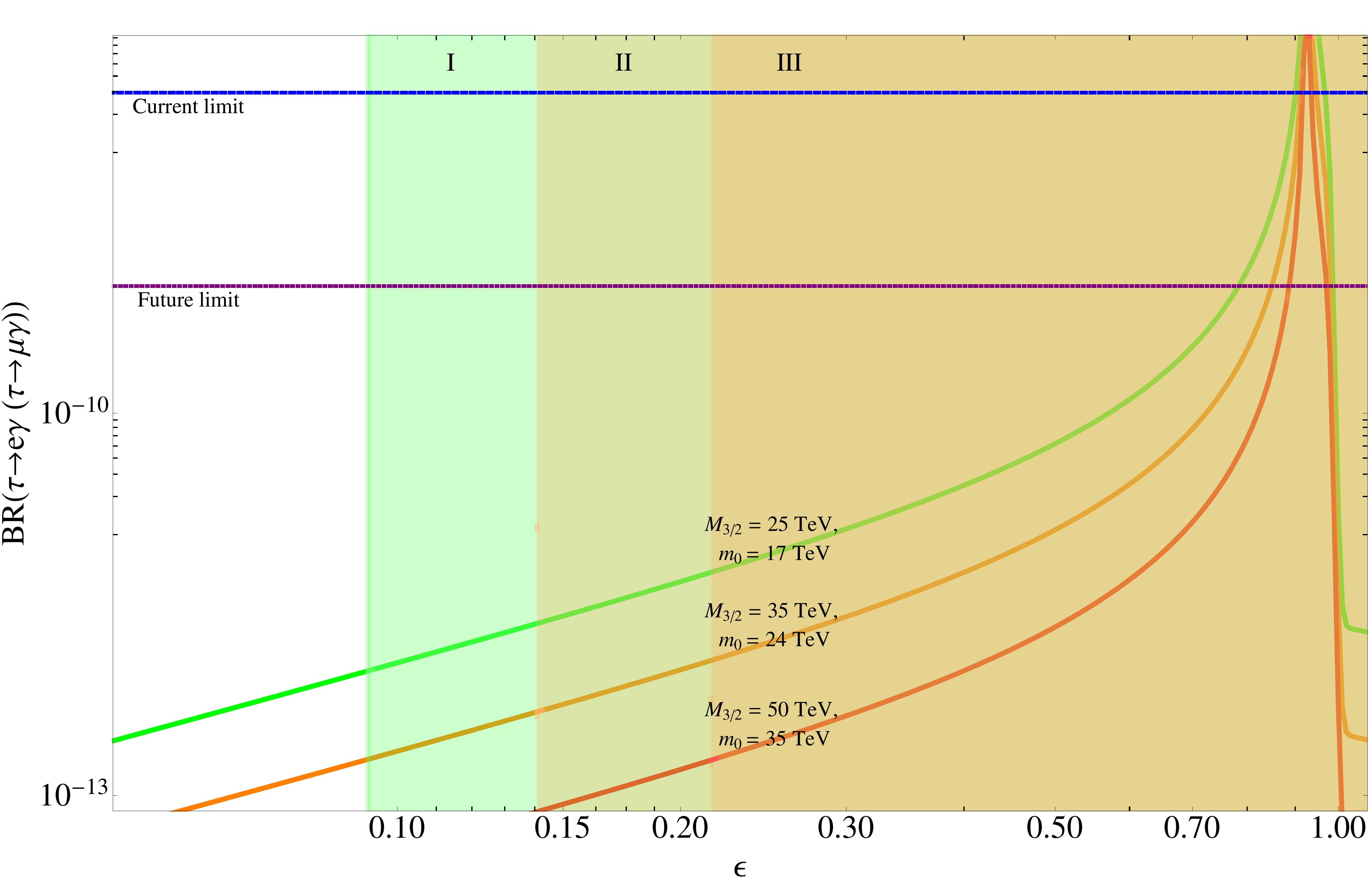}
\caption{Branching ratio of $\tau \to e (\mu) \gamma$ as a function of the small parameter $\epsilon$ for textures 3 (4). The curves represent $M_{3/2} = 25,~35,~50$ TeV for green, orange and pink respectively. The dotted blue line represents the current experimental limit, and the dotted purple line is the future sensitivity. We fix $\mu = 1.4$ TeV, $m_{\tilde{g}}=1.5$ TeV, $M_1 = 400$ GeV, $M_2 = 600$ GeV and $\tan\beta = 7$.}
\label{taus.FIG}
\end{figure}

The third and fourth textures that we consider will clearly give no contribution to the decay of $\mu \to e \gamma$. They give rise only to non-zero $\tau \to e \gamma$ and $\tau \to \mu \gamma$ decays for the textures 3 and 4 respectively. Since the $\epsilon$ dependence is the same in both cases, we need only show one plot, which applies to both cases. This is shown in fig. \ref{taus.FIG}.

Quite clearly, both the current and future sensitivities are already excluded in principle by $\Delta m_K$. However, it is worth noting that if $\tau \to e (\mu) \gamma$ were to be observed, this would imply that the constraints from $\Delta m_K$ do not apply to the leptonic sector, which would in turn tell us that the K\"ahler potential either knows the difference between quarks and leptons, or that there may be some mechanism making the squarks heavier than the sleptons (although that is not expected in the M-theory compactification).

\subsection{$\mu \to e$ conversion in Aluminium}

While the process of $\mu \to e$ conversion in nuclei is suppressed relative to the rare decay of $\mu \to e \gamma$, the future sensitivity of the conversion process in Aluminium promises to probe further than future $\mu \to e \gamma$ experiments (c.f. Table \ref{Limits.TAB}). Therefore, we present here the numerical results for the calculation of the conversion process in Aluminium in the mass insertion approximation. As explained earlier, while there is some deviation of order a few percent between the mass insertion approximation and the eigenstate calculation for larger $\epsilon$, since $\mu \to e$ conversion probes the region of small $\epsilon$, using the mass insertion approximation is justified. 

\begin{figure}[H]
\centering
\includegraphics[scale=0.35]{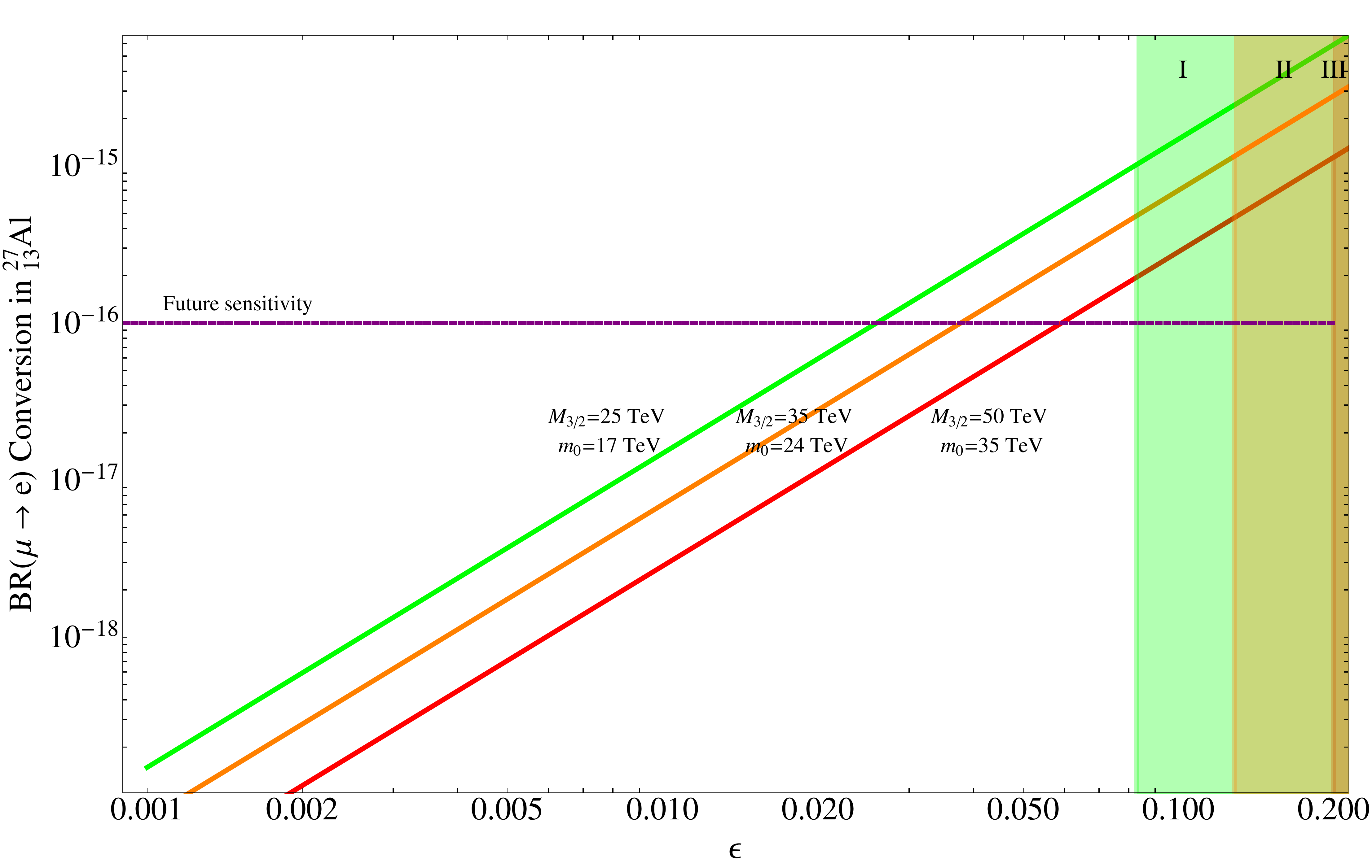}
\caption{The branching ratio for $\mu \to e$ conversion in Aluminium as a function of the small parameter $\epsilon$. The green, orange and red lines are for $M_{3/2} = 25,~35,~50$ TeV respectively, corresponding to $m_0 = 17,~24,~35$ TeV. The shaded regions I, II and III denote the exclusions due to $\Delta{m_K}$ constraints for $M_{3/2} = 25,~35,~50$ TeV respectively. The dotted purple line is the future sensitivity. We fix $\mu = 1.4$ TeV, $m_{\tilde{g}}=1.5$ TeV, $M_1 = 400$ GeV, $M_2 = 600$ GeV and $\tan\beta = 7$.}
\label{Conversion.FIG}
\end{figure}

We see from Fig. \ref{Conversion.FIG} that the future sensitivity of $\Order(10^{-16})$ to muon conversion in Aluminium will prove to be an improvement over $\Delta m_K$ of order a few, and a slight improvement over BR($\mu \to e \gamma$). The future sensitivity should be able to impose bounds of $\epsilon \lesssim 0.026,~0.038,~0.06$ for $M_{3/2} = 25,~35,~50$ ($m_0 = 17,~24,~35$) TeV respectively\footnote{Note that the bounds for conversion do not quite scale the same way as for BR($\mu \to e \gamma$). This is due to interplay between the dipole and the photonic penguin operators.}. If the sensitivity could be improved to $\Order(10^{-17})$, the bounds would be improved to $\epsilon \lesssim 0.008,~0.011,~0.019$, which would be a very clear improvement over all other current and future experimental constraints.

\subsection{Summary of constraints, current and future}

Here we present for convenience a summary of the constraints from all processes, currently and in the future.\footnote{We present for reference the calculated sensitivity from $\mu \to 3e$ as forecast in \cite{Blondel:2013ia}, which, if realised, is potentially comparable with other observables.}

\begin{table}[H]
\centering
\begin{tabular}{c c c c}
\hline
\textbf{Process} & \textbf{Scalar mass $m_0$ (TeV)}& \textbf{Current bound ($\epsilon$)} & \textbf{Future sensitivity ($\epsilon$)} \\
\hline
$\Delta m_K$& 17 & 0.09  & -  \\
& 24 & 0.14 & - \\
& 35 & 0.22  & - \\
\hline
BR($\mu \to e \gamma$) Texture 1 &17 & 0.06  & 0.02  \\
& 24 & 0.14 & 0.06 \\
& 35 & 0.24 & 0.11 \\
\hline
BR($\mu \to e \gamma$) Texture 2 &17 & 0.08  & 0.03  \\
& 24 & 0.19 & 0.06 \\
& 35 & 0.5 & 0.15 \\
\hline
BR($\tau \to e (\mu) \gamma$) Texture 3 (4) &17 & 0.9  & 0.8  \\
& 24 & 0.9 & 0.85 \\
& 35 & 0.9 & 0.9 \\
\hline
BR($\mu \to e $)$_{^{27}_{13}\text{Al}}$  &17 & -  & 0.026  \\
& 24 & - & 0.038 \\
& 35 & - & 0.06 \\
\hline
BR($\mu \to 3e $)  &17 & -  & 0.031  \\
& 24 & - & 0.07 \\
& 35 & - & 0.18 \\
\hline
\end{tabular}
\caption{Limits on $\epsilon$ from all flavour observables considered, present and future. Note that the current constraint from $\mu \to e \gamma$ is comparable with that from $\Delta m_K$, and in the future will be the stronger of the two.}
\label{EpsLimits.TAB}
\end{table}




\section{Conclusion}
\label{Conc.SEC}

The flavour structure of the K\"ahler potential in supergravity theories has potentially important effects on low-energy observables. The current stringent limits on lepton-flavour violating processes, in particular the branching ratio $\mu \to e \gamma$ being so small, sets the bar for gravity-mediated supersymmetry theories to meet. We have studied a generic K\"ahler potential containing terms mixing the visible sector with a hidden sector in the context of a compactified M-theory. The higher-order corrections to this potential that couple the two sectors can have important implications for low-energy flavour observables. While $U(1)$ symmetries or geometrical effects may protect the K\"ahler potential from having flavour off-diagonal terms at high scales, we consider the possibility that these are broken at some scale below the Planck mass, resulting in suppressed flavour violation. 

We make a parameterisation of the higher-order correction term coupling the visible sector to the hidden sector, which enables us to analyse how flavour observables can be used to probe the flavour structure of the K\"ahler potential. We find that the limits on lepton-flavour violating decays translate into a strong bound on off-diagonal terms in the scalar mass matrix, parameterised by $\epsilon$. For a gravitino mass of order $25-50$ TeV ($m_0 =17-35$ TeV), as implied in the compactified M-theory case, the current experimental bound on BR$(\mu \to e \gamma$) already sets bounds on $\epsilon$ of between $6 \times10^{-2}$  and $\sim 2 \times 10^{-1}$. This is already more constraining than $\Delta m_K$ for $M_{3/2} \lesssim 35$ TeV. The future experimental bounds promise to improve these bounds by a factor of two to three, and improve on the constraints from $\Delta m_K$ for all the values of $M_{3/2}$ considered here. Future experiments attempting to measure $\mu \to e$ conversion in Aluminium will further improve the bounds. The expectation is that BR($\tau \to e (\mu) \gamma$) should not be seen, as it is already ruled out by $\Delta m_K$, although if it were to be seen, it would suggest some non-trivial interplay between the quark and lepton sectors in the K\"ahler potential. One could argue that, given the already fairly strong contraints, future measurements of $\mu \to e \gamma$ or $\mu \to e$ conversion in Aluminium will either find a non-zero branching ratio due to K\"ahler potential effects in the not too distant future, or else the K\"ahler potential is flavour-diagonal, or nearly so. Such a flavour-diagonal K\"ahler potential has not been anticipated.

We see that LFV processes have the potential to probe aspects of M-theory compactifications that pose interesting conceptual and technical challenges for model-builders and for the underlying theory. Additionally, we note that LFV processes have the potential to become the strongest constraints on heavy supersymmetric scalars with generic flavour structure, surpassing the constraints from $\Delta m_K$, and will thus be an important probe of models with a mini-split spectrum. The current experimental limits suggest that, for the preferred gravitino masses \cite{Acharya:2007rc, Acharya:2008hi,Acharya:2008zi, Acharya:2012tw}, the K\"ahler potential must already be surprisingly flavour-diagonal. This would have to be explained either by symmetry arguments, or by geometrical arguments from the underlying theory. Thus, improvement in experimental sensitivity in $\mu \to e \gamma$ and $\mu \to e$ conversion should either give a non-zero signal, or point towards the surprising result that the K\"ahler potential is flavour-diagonal, or nearly so.

\section*{Acknowledgements}

We would like to thank John Ellis, James Wells and Bob Zheng for useful discussions during the course of this work. A special thanks to Aaron Pierce for his advice on the constraints from $\Delta m_K$, and for technical discussions. SARE and GLK are supported in part by Department of Energy grant DE-SC0007859.

\appendix

\section{Calculation of $\Gamma_{\alpha\beta}$}
\label{ScalarsCalc.APP}
In this section we go through the explicit calculation of $\Gamma_{\alpha\beta}$. Given the K\"ahler potential in equation (\ref{KP.EQ}) above, we can see that the K\"ahler metric $\tilde{K}_{\alpha\beta}$ is given by
\beq
\label{Metric.EQ}
\tilde{K}_{\alpha\beta} = \frac{\kappa_{\alpha\beta}}{V_7} + \frac{c_{\alpha\beta} \bar{\phi}\phi}{3V_7^2}
\eeq
and the inverse K\"ahler metric is then approximately
\beq
\label{InvMetric.EQ}
\tilde{K}^{\alpha\beta} \approx \kappa^{\alpha\beta} V_7 - \frac{c^{\alpha\beta} \bar{\phi}\phi}{3}
\eeq 

In the course of the calculation, we use the following identities from \cite{Acharya:2008hi}
\begin{align}
&\sum_{i=1}^N s_i \hat{K}_i = -7,~~~\hat{K}_i \equiv \frac{\partial \hat{K}}{\partial s_i},~~~\hat{K} = -3\log(4\pi^{1/3}V_7)\\
&\sum_{i=1}^N s_i \frac{\partial \kappa_{\alpha\beta}(s_i)}{\partial s_i} = \lambda \kappa_{\alpha\beta} (s_i),~~~\frac{\partial}{\partial s_i}\frac{1}{V_7} = \frac{\hat{K}_i}{3V_7}
\end{align}
and the assumption, explained in section \ref{Scalars.SSEC}, that we may decompose $c_{\alpha\beta}$ such that
\beq
c_{\alpha\beta} = c \cdot \kappa_{\alpha\delta}\sigma^\delta_\beta
\eeq
where $\sigma^\delta_\beta$ is some unknown matrix that is assumed not to depend on the moduli fields and c is the same factor from \cite{Acharya:2008hi} which can take on a value between 0 and 1.

We also use the results from \cite{Acharya:2008hi} 
\begin{align}
e^{\hat{K}/2} F^i &\simeq 2 i s_i m_{1/2}^{tree} \\
e^{\hat{K}/2} F^\phi &\simeq \phi \left(\frac{7}{3}-P_{eff}\right) m_{1/2}^{tree}
\end{align}
where $P_{eff}$ is a constant defined in Eqs. (125), (141) of \cite{Acharya:2008hi}, and is a constant. It depends on the terms $A_1$ and $A_2$ of the superpotential, the hidden sector field value $\phi_0$, as well as the hidden sector gauge groups which are $SU(P+1)$ and $SU(Q)$. We replicate here the expressions in \cite{Acharya:2008hi}
\beq
P_{eff} \equiv P\ln\left( \frac{Q A_1 \phi_0^2}{P A_2}\right) \simeq \frac{14(3(Q-P) -2)}{3(3(Q-P) - 2\sqrt{6(Q-P)}}
\eeq

We then break up the expression for $\Gamma_{\alpha\beta}$ into two more easily tractable pieces, $A_{\alpha\beta}$: $e^{\hat{K}}\bar{F}^{\bar{m}}(\partial_{\bar{m}} \partial_n \tilde{K}_{\alpha\beta})F^n$ and $B_{\alpha\beta}$: $e^{\hat{K}}\bar{F}^{\bar{m}}(\partial_{\bar{m}} \tilde{K}_{\alpha\gamma}\tilde{K}^{\gamma\delta}\partial_n \tilde{K}_{\delta\beta})F^n$ such that $\Gamma_{\alpha\beta} = A_{\alpha\beta} - B_{\alpha\beta}$.

Starting first with term $A_{\alpha\beta}$, we find that it can be written in the following way:
\begin{align}
\nonumber e^{\hat{K}}\bar{F}^{\bar{m}}(\partial_{\bar{m}} \partial_n \tilde{K}_{\alpha\beta})F^n &= e^{\hat{K}}\bar{F}^{\bar{j}}(\partial_{\bar{j}} \partial_i \tilde{K}_{\alpha\beta})F^i\\
\nonumber &+ e^{\hat{K}}\bar{F}^{\bar{j}}(\partial_{\bar{j}} \partial_\phi \tilde{K}_{\alpha\beta})F^\phi\\
\nonumber &+ e^{\hat{K}}\bar{F}^{\bar{\phi}}(\partial_{\bar{\phi}} \partial_i \tilde{K}_{\alpha\beta})F^i\\
&+ e^{\hat{K}}\bar{F}^{\bar{\phi}}(\partial_{\bar{\phi}} \partial_\phi \tilde{K}_{\alpha\beta})F^\phi
\end{align}
so that $I$: $e^{\hat{K}}\bar{F}^{\bar{j}}(\partial_{\bar{j}} \partial_i \tilde{K}_{\alpha\beta})F^i$, $II$: $e^{\hat{K}}\bar{F}^{\bar{j}}(\partial_{\bar{j}} \partial_\phi \tilde{K}_{\alpha\beta})F^\phi$, $III$:, $e^{\hat{K}}\bar{F}^{\bar{\phi}}(\partial_{\bar{\phi}} \partial_i \tilde{K}_{\alpha\beta})F^i$ and $IV$: $e^{\hat{K}}\bar{F}^{\bar{\phi}}(\partial_{\bar{\phi}} \partial_\phi \tilde{K}_{\alpha\beta})F^\phi$.

Term $I$ is given by
\begin{align}
e^{\hat{K}}\bar{F}^{\bar{j}}(\partial_{\bar{j}} \partial_i \tilde{K}_{\alpha\beta})F^i &= e^{\hat{K}}\bar{F}^{\bar{j}}\left(\partial_{\bar{j}}\left( \frac{\partial_i \kappa_{\alpha\beta} }{V_7} + \frac{\kappa_{\alpha\beta} \hat{K}_i}{3V_7} + \frac{\partial_i \kappa_{\alpha\gamma} \sigma^\gamma_\beta c \bar{\phi}\phi}{3V_7^2} + \frac{2c_{\alpha\beta} \bar{\phi}\phi \hat{K}_i }{9V_7^2} \right)\right)F^i 
\\
\nonumber&= e^{\hat{K}}\bar{F}^{\bar{j}}\Bigg( \frac{\partial_{\bar{j}} \partial_i \kappa_{\alpha\beta}}{V_7} + \frac{\partial_i \cab \hat{K}_i}{3V_7} + \frac{\partial_{\bar{j}}\kab \hat{K}_i}{3V_7} + \frac{\kab \hat{K}_{\bar{j}} \hat{K}_i}{9V_7} \\
&+ \frac{\partial_{\bar{j}} \partial_i \kappa_{\alpha\gamma}\sigma^\gamma_\beta c\bar{\phi}\phi}{3V_7^2} + \frac{2 \partial_i \kappa_{\alpha\gamma}\sigma^\gamma_\beta c\bar{\phi}\phi \hat{K}_{\bar{j}}}{9V_7^2} + \frac{2 \partial_{\bar{j}} \kappa_{\alpha\gamma}\sigma^\gamma_\beta c\bar{\phi}\phi \hat{K}_i}{9V_7^2} + \frac{4 \cab \bar{\phi}\phi \hat{K}_{\bar{j}} \hat{K}_i}{27V_7^2} \Bigg) F^i \\
&= 4\left(m_{1/2}^{tree}\right)^2 \left[ \left(\lambda - \frac{7}{3}\right)^2 \frac{\kab}{V_7} + \left(\lambda - \frac{14}{3}\right)^2 \frac{\cab \bar{\phi}\phi}{3V_7^2} \right]
\end{align}
then terms $II$ and $III$ cancel, leaving term $IV$, which is
\begin{align}
e^{\hat{K}}\bar{F}^{\bar{\phi}}(\partial_{\bar{\phi}} \partial_\phi \tilde{K}_{\alpha\beta})F^\phi &= e^{\hat{K}}\bar{F}^{\bar{\phi}}\left(\partial_{\bar{\phi}} \partial_\phi \frac{\cab}{3V_7^2}\right) F^\phi \\
&= \left(\frac{7}{3}-P_{eff} \right)^2 \left(m_{1/2}^{tree}\right)^2 \frac{\cab \bar{\phi}\phi}{3V_7^2}
\end{align}
such that term $A_{\alpha\beta}$ is just
\beq
A_{\alpha\beta} =  \left(m_{1/2}^{tree}\right)^2 \left[ 4\left( \left(\lambda - \frac{7}{3}\right)^2 \frac{\kab}{V_7} + \left(\lambda - \frac{14}{3}\right)^2 \frac{\cab \bar{\phi}\phi}{3V_7^2}\right) + \left(\frac{7}{3}-P_{eff} \right)^2 \frac{\cab \bar{\phi}\phi}{3V_7^2} \right]
\eeq

Moving on to the second term $B_{\alpha\beta}$, we write
\begin{align}
B_{\alpha\beta} &= e^{\hat{K}}\bar{F}^{\bar{m}}(\partial_{\bar{m}} \tilde{K}_{\alpha\gamma}\tilde{K}^{\gamma\delta}\partial_n \tilde{K}_{\delta\beta})F^n \\
&= e^{\hat{K}}\bar{F}^{\bar{m}}\left(\partial_{\bar{m}} \tilde{K}_{\alpha\gamma}\tilde{K}^{\gamma\delta} \left( \left( \frac{\partial_i \kappa_{\delta\beta} }{V_7} + \frac{\kappa_{\delta\beta} \hat{K}_i}{3V_7} + \frac{\partial_i \kappa_{\delta\xi} \sigma^\xi_\beta c\bar{\phi}\phi}{3V_7^2} + \frac{2c_{\delta\beta} \bar{\phi}\phi \hat{K}_i }{9V_7^2} \right)F^i +\left(\frac{c_{\delta\beta} \bar{\phi}}{3V_7^2}\right) F^\phi \right)\right)\\
\nonumber&=  \left(m_{1/2}^{tree}\right) e^{\hat{K}/2}\bar{F}^{\bar{m}}\Bigg(\partial_{\bar{m}} \tilde{K}_{\alpha\gamma}\Bigg( 2i \left[\left(\lambda - \frac{7}{3}\right)\left(\delta^\gamma_\beta - \frac{c^{\gamma\delta}\kappa_{\delta\beta} \bar{\phi}\phi}{3V_7}\right) + \left(\lambda - \frac{14}{3}\right)\left(\frac{\kappa^{\gamma\delta}c_{\delta\beta} \bar{\phi}\phi}{3V_7} - \frac{c^{\gamma\delta}c_{\delta\beta} (\bar{\phi}\phi)^2}{9V_7^2}\right)\right]\\
&\hspace{7mm}+ \left(\frac{7}{3} - P_{eff}\right)\left[\frac{\kappa^{\gamma\delta}c_{\delta\beta} \bar{\phi}\phi}{3V_7} - \frac{c^{\gamma\delta}c_{\delta\beta} (\bar{\phi}\phi)^2}{9V_7^2}\right]\Bigg)\Bigg) \\
\nonumber&= \left(m_{1/2}^{tree}\right)^2 \Bigg\{ (-2i) \left[\left(\lambda - \frac{7}{3}\right) \frac{\kappa_{\alpha\gamma}}{V_7} + \left(\lambda - \frac{14}{3}\right) \frac{c_{\alpha\gamma} \bar{\phi}\phi}{3V_7^2}\right] + \left( \frac{7}{3} - P_{eff}\right) \frac{c_{\alpha\gamma} \bar{\phi}\phi}{3V_7^2}\Bigg\} \\
\nonumber& \hspace{7mm}\times \Bigg\{  2i \left[\left(\lambda - \frac{7}{3}\right)\left(\delta^\gamma_\beta - \frac{c^{\gamma\delta}\kappa_{\delta\beta} \bar{\phi}\phi}{3V_7}\right) + \left(\lambda - \frac{14}{3}\right)\left(\frac{\kappa^{\gamma\delta}c_{\delta\beta} \bar{\phi}\phi}{3V_7} - \frac{c^{\gamma\delta}c_{\delta\beta} (\bar{\phi}\phi)^2}{9V_7^2}\right)\right] \\
& \hspace{13mm}+  \left(\frac{7}{3} - P_{eff}\right)\left[\frac{\kappa^{\gamma\delta}c_{\delta\beta} \bar{\phi}\phi}{3V_7} - \frac{c^{\gamma\delta}c_{\delta\beta} (\bar{\phi}\phi)^2}{9V_7^2}\right]\Bigg\} \\
\nonumber &=\left(m_{1/2}^{tree}\right)^2 \Bigg\{ 4 \bigg[ \left(\lambda - \frac{7}{3}\right)^2\left(\frac{\kab}{V_7} - \frac{\kappa_{\alpha\gamma}c^{\gamma\delta}\kappa_{\delta\beta} \bar{\phi}\phi}{3V_7^2}\right) \\ 
\nonumber&\hspace{30mm}+ \left(\lambda - \frac{7}{3}\right)\left(\lambda - \frac{14}{3}\right)\left(\frac{\kappa_{\alpha\gamma}\kappa^{\gamma\delta}c_{\delta\beta} \bar{\phi}\phi}{3V_7^2} - \frac{\kappa_{\alpha\gamma}c^{\gamma\delta}c_{\delta\beta} (\bar{\phi}\phi)^2}{9V_7^3}\right)\\
\nonumber&\hspace{30mm}+ \left(\lambda - \frac{14}{3}\right)\left(\lambda - \frac{7}{3}\right)\left( \frac{c_{\alpha\beta} \bar{\phi}\phi}{3V_7^2} - \frac{c_{\alpha\gamma} c^{\gamma\delta} \kappa_{\delta\beta} (\bar{\phi}\phi)^2}{9V_7^3}\right)\\ 
\nonumber&\hspace{30mm}+ \left(\lambda - \frac{14}{3}\right)^2 \left(\frac{c_{\alpha\gamma} \kappa^{\gamma\delta} c_{\delta\beta} (\bar{\phi}\phi)^2}{9V_7^3} - \frac{c_{\alpha\gamma} c^{\gamma\delta} c_{\delta\beta} (\bar{\phi}\phi)^3}{27V_7^4}\right)\bigg] \\
\nonumber&\hspace{20mm}-2i\left(\frac{7}{3} - P_{eff}\right) \bigg[ \left(\lambda - \frac{7}{3}\right)\left( \frac{\kappa_{\alpha\gamma}\kappa^{\gamma\delta}c_{\delta\beta} \bar{\phi}\phi}{3V_7^2} - \frac{\kappa_{\alpha\gamma}c^{\gamma\delta}c_{\delta\beta} (\bar{\phi}\phi)^2}{9V_7^3} - \frac{\cab \bar{\phi}\phi}{3V_7^2} + \frac{c_{\alpha\gamma}c^{\gamma\delta}\kappa_{\delta\beta} (\bar{\phi}\phi)^2}{9V_7^3}\right) \\ 
\nonumber&\hspace{30mm}+ \left(\lambda - \frac{14}{3}\right) \left(\frac{c_{\alpha\gamma}\kappa^{\gamma\delta}c_{\delta\beta} (\bar{\phi}\phi)^2}{9V_7^3} - \frac{c_{\alpha\gamma}c^{\gamma\delta}c_{\delta\beta} (\bar{\phi}\phi)^3}{27V_7^4} - \frac{c_{\alpha\gamma}\kappa^{\gamma\delta}c_{\delta\beta} (\bar{\phi}\phi)^2}{9V_7^3} + \frac{c_{\alpha\gamma}c^{\gamma\delta}c_{\delta\beta} (\bar{\phi}\phi)^3}{27V_7^4}\right)\bigg]\\
& \hspace{20mm}+\left(\frac{7}{3} - P_{eff}\right)^2 \left[ \frac{c_{\alpha\gamma}\kappa^{\gamma\delta}c_{\delta\beta} (\bar{\phi}\phi)^2}{9V_7^3} - \frac{c_{\alpha\gamma}c^{\gamma\delta}c_{\delta\beta} (\bar{\phi}\phi)^3}{27V_7^3}\right] \Bigg\}
\end{align}
We can then use 
\beq
\kappa_{\alpha\delta}\kappa^{\delta\gamma} = \delta_\alpha^\gamma
\eeq
and
\beq
\kappa_{\alpha\delta}\sigma^\delta_\beta = \sigma_\alpha^\delta \kappa_{\delta\beta}
\eeq
to see that the complex entry is just zero, such that 
\begin{align}
\nonumber B_{\alpha\beta}&=\left(m_{1/2}^{tree}\right)^2 \Bigg\{ 4 \bigg[ \left(\lambda - \frac{7}{3}\right)^2\left(\frac{\kab}{V_7} - \frac{\kappa_{\alpha\gamma}c^{\gamma\delta}\kappa_{\delta\beta} \bar{\phi}\phi}{3V_7^2}\right) \\ 
\nonumber&\hspace{30mm}+ \left(\lambda - \frac{7}{3}\right)\left(\lambda - \frac{14}{3}\right)\left(\frac{\kappa_{\alpha\gamma}\kappa^{\gamma\delta}c_{\delta\beta} \bar{\phi}\phi}{3V_7^2} - \frac{\kappa_{\alpha\gamma}c^{\gamma\delta}c_{\delta\beta} (\bar{\phi}\phi)^2}{9V_7^3}\right)\\
\nonumber&\hspace{30mm}+ \left(\lambda - \frac{14}{3}\right)\left(\lambda - \frac{7}{3}\right)\left( \frac{c_{\alpha\beta} \bar{\phi}\phi}{3V_7^2} - \frac{c_{\alpha\gamma} c^{\gamma\delta} \kappa_{\delta\beta} (\bar{\phi}\phi)^2}{9V_7^3}\right)\\ 
\nonumber&\hspace{30mm}+ \left(\lambda - \frac{14}{3}\right)^2 \left(\frac{c_{\alpha\gamma} \kappa^{\gamma\delta} c_{\delta\beta} (\bar{\phi}\phi)^2}{9V_7^3} - \frac{c_{\alpha\gamma} c^{\gamma\delta} c_{\delta\beta} (\bar{\phi}\phi)^3}{27V_7^4}\right)\bigg] \\
& \hspace{20mm}+\left(\frac{7}{3} - P_{eff}\right)^2 \left[ \frac{c_{\alpha\gamma}\kappa^{\gamma\delta}c_{\delta\beta} (\bar{\phi}\phi)^2}{9V_7^3} - \frac{c_{\alpha\gamma}c^{\gamma\delta}c_{\delta\beta} (\bar{\phi}\phi)^3}{27V_7^3}\right] \Bigg\}
\end{align}

Putting the pieces together finally, we have that 
\begin{align}
\nonumber \Gamma_{\alpha\beta} &= \left(m_{1/2}^{tree}\right)^2 \Bigg\{ 4 \left[ \frac{49}{9} \frac{\cab \bar{\phi}\phi}{3V_7^2} + \lambda\left(\lambda- \frac{14}{3}\right) \frac{\sigma_\alpha^\delta c_{\delta\beta} c(\bar{\phi}\phi)^2}{9V_7^3} + \left(\lambda- \frac{14}{3}\right)^2 \frac{\sigma^\gamma_\alpha \sigma_\gamma^\delta c_{\delta\beta} c^2(\bar{\phi}\phi)^3}{27V_7^4}\right] \\
&\hspace{25mm}+ \left(\frac{7}{3} - P_{eff}\right)^2 \left[ \frac{\cab\bar{\phi}\phi}{3V_7^2} - \frac{\sigma_\alpha^\delta c_{\delta\beta} c(\bar{\phi}\phi)^2}{9V_7^3} + \frac{\sigma^\gamma_\alpha \sigma_\gamma^\delta c_{\delta\beta} c^2(\bar{\phi}\phi)^3}{27V_7^4}\right] \Bigg\}
\end{align}

\section{Calculation of the Trilinears}
\label{TrilinearsCalc.APP}
We start from Eq. (\ref{Trilinears.EQ}) and then apply similar conditions as above to derive a useful form of the Trilinears. We may break up Eq. (\ref{Trilinears.EQ}) into 3 parts,
\begin{align}
I) &:~\frac{\overline{W}}{|W|}e^{K/2}F^mK_m Y'_{\alpha\beta\gamma}\\
II) &:~\frac{\overline{W}}{|W|}e^{K/2}F^m \partial_m Y'_{\alpha\beta\gamma}\\
III) &:~\frac{\overline{W}}{|W|}e^{K/2}F^m \left[\tilde{K}^{\delta\rho}\partial_m \tilde{K}_{\rho\alpha} Y'_{\delta\beta\gamma} + \alpha \leftrightarrow \beta + \alpha \leftrightarrow \gamma \right]
\end{align}

Starting with part $I)$, we may use the properties set out above to find
\beq
\frac{\overline{W}}{|W|}e^{K/2}F^mK_m Y'_{\alpha\beta\gamma} = \left( \frac{\phi_0^2}{V_7} + \frac{7}{P_{eff}}\right)\left(1+ \frac{2V_7}{(Q-P)\phi_0^2}\right)m_{3/2}Y'_{\alpha\beta\gamma}
\eeq
as is also given in \cite{Acharya:2008hi}. Since there is no dependence on $\tilde{K}_{\alpha\beta}$, $II)$ is also the same as in \cite{Acharya:2008hi}, and is
\beq
\frac{\overline{W}}{|W|}e^{K/2}F^m \partial_m Y'_{\alpha\beta\gamma} =\frac{4\pi}{P_{eff}} \left(1+ \frac{2V_7}{(Q-P)\phi_0^2}\right)V_{Q^{\alpha\beta\gamma}}m_{3/2}Y'_{\alpha\beta\gamma}
\eeq
Part $III)$ however does have dependence on the metric $\tilde{K}_{\alpha\beta}$. We therefore calculate the form of the relevant part of $III)$. Since the three terms in $III)$ are equivalent up to interchanging of indices, it is sufficient to calculate one of them to infer the rest.
\begin{align}
e^{K/2}F^m \tilde{K}^{\delta\rho}\partial_m \tilde{K}_{\rho\alpha} = e^{K/2} \left(F^i \tilde{K}^{\delta\rho}\partial_i \tilde{K}_{\rho\alpha} + F^{\phi} \tilde{K}^{\delta\rho}\partial_{\phi} \tilde{K}_{\rho\alpha} \right)
\end{align}
Turning our attention first to the part dependent on the moduli fields, we use that 
\beq
\partial_i \tilde{K}_{\rho\alpha} = \frac{\partial_i \kappa_{\rho\alpha}}{V_7} + \frac{\kappa_{\rho\alpha}\hat{K}_i}{3V_7} + \frac{c \partial_i \kappa_{\rho\delta}\sigma^\delta_\alpha \overline{\phi}\phi}{3V_7^2} + \frac{2c\kappa_{\rho\delta}\sigma^\delta_\alpha  \overline{\phi}\phi\hat{K}_i}{9V_7^2}
\eeq
therefore using the expression for the inverse K\"ahler metric in Eq. (\ref{InvMetric.EQ}) we find
\beq
e^{K/2} F^i \tilde{K}^{\delta\rho}\partial_i \tilde{K}_{\rho\alpha} = \frac{-1}{P_{eff}}\left(1+ \frac{2V_7}{(Q-P)\phi_0^2}\right)\left[ \left(\lambda-\frac{7}{3}\right) \delta^\delta_\alpha - \left(\lambda-\frac{14}{3}\right) \frac{c^2\sigma^\delta_\sigma \sigma^\sigma_\alpha (\overline{\phi}\phi)^2}{9V_7^2} \right] m_{3/2}
\eeq
Similarly, we calculate the part dependent on the hidden sector fields, using
\beq
\partial_\phi \tilde{K}_{\rho\alpha} = \frac{c\kappa_{\rho\sigma}\sigma^\sigma_\alpha \overline{\phi}}{3V_7^2}
\eeq
we find that
\beq
 e^{K/2}F^{\phi} \tilde{K}^{\delta\rho}\partial_{\phi} \tilde{K}_{\rho\alpha} = \left(1 - \frac{7}{3P_{eff}}\right) \left(1+ \frac{2V_7}{(Q-P)\phi_0^2}\right) \left[ \frac{c\sigma^\delta_\alpha \overline{\phi}\phi}{3V_7^2} - \frac{c^2\sigma^\delta_\sigma \sigma^\sigma_\alpha (\overline{\phi}\phi)^2}{9V_7^2}\right] m_{3/2}
\eeq
Putting the two parts together, and then combining with $I)$ and $II)$, we may write that
\begin{align}
\nonumber A'_{\alpha\beta\gamma} = \frac{\overline{W}}{|W|} \Bigg\{ &\left[ \left(\frac{\phi_0^2}{V_7} + \frac{7}{P_{eff}}\right) + \frac{4\pi V_{Q^{\alpha\beta\gamma}}}{P_{eff}}\right] Y'_{\alpha\beta\gamma} \\
\nonumber &- \Bigg[ \left(1 - \frac{7}{3P_{eff}}\right)  \left( \frac{c\sigma^\delta_\alpha \overline{\phi}\phi}{3V_7^2} - \frac{c^2\sigma^\delta_\sigma \sigma^\sigma_\alpha (\overline{\phi}\phi)^2}{9V_7^2}\right) \\
\nonumber &- \frac{1}{P_{eff}}\left( \left(\lambda-\frac{7}{3}\right) \delta^\delta_\alpha - \left(\lambda-\frac{14}{3}\right) \frac{c^2\sigma^\delta_\sigma \sigma^\sigma_\alpha (\overline{\phi}\phi)^2}{9V_7^2} \right)\Bigg] Y'_{\delta\beta\gamma} \\
&+ \alpha \leftrightarrow \beta + \alpha \leftrightarrow \gamma \Bigg\}\left(1+ \frac{2V_7}{(Q-P)\phi_0^2}\right) m_{3/2}
\end{align}

We then use the definition of the physical Yukawa couplings, and disregard terms higher-order in $\phi_0^2/V_7$ to write
\begin{align}
\nonumber A'_{\alpha\beta\gamma} &\simeq \Bigg\{ \left[ \frac{\phi_0^2}{V_7}\left(Y_{\alpha\beta\gamma} - c\left(1-\frac{7}{3P_{eff}}\right)\left[\frac{\sigma^\delta_\alpha}{3} Y_{\delta\beta\gamma} + \alpha\leftrightarrow\beta + \alpha\leftrightarrow\gamma \right]\right)\right] \\&+ \frac{4\pi V_{Q^{\alpha\beta\gamma}} + 3\lambda}{P_{eff}}Y_{\alpha\beta\gamma}\Bigg\}\left(1+ \frac{2V_7}{(Q-P)\phi_0^2}\right) m_{3/2}
\end{align}

\section{Expressions for LFV processes}
\label{LFVprocesses.APP}

\subsection{Expressions for $l_i \to l_j \gamma$}

The amplitude for the process $l_i \rightarrow l_j \gamma$ is given by \cite{Hisano:1995cp, Casas:2001sr}:
\begin{align}
&T = \epsilon^\alpha \bar{l}_j m_{l_i} i \sigma_{\alpha\beta} q^\beta (A_L P_L +A_R P_R)l_i,
\\ \nonumber
\\
&A_{L,R} = A_{L,R}^{(c)} + A_{L,R}^{(n)},
\\ \nonumber
\\
&A_{L,R}^{(c)} = - \frac{e}{16\pi^2}\frac{1}{m^2_{\tilde{\nu}_X}}\left[ C^{L,R}_{jAX} {C^{L,R}_{iAX}}^* I_1(M_{\tilde{\chi}^-_A}/m^2_{\tilde{\nu}_X}) + C^{L,R}_{jAX} {C^{R,L}_{iAX}}^* \frac{M_{\tilde{\chi}^-_A}}{m_{l_i}} I_2(M_{\tilde{\chi}^-_A}/m^2_{\tilde{\nu}_X}) + \right]
\\ \nonumber
\\
&A_{L,R}^{(n)} =  \frac{e}{16\pi^2}\frac{1}{m^2_{\tilde{l}_X}}\left[ N^{L,R}_{jAX} {N^{L,R}_{iAX}}^* J_1(M_{\tilde{\chi}^0_A}/m^2_{\tilde{l}_X}) + N^{L,R}_{jAX} {N^{R,L}_{iAX}}^* \frac{M_{\tilde{\chi}^0_A}}{m_{l_i}} J_2(M_{\tilde{\chi}^0_A}/m^2_{\tilde{l}_X}) + \right]
\end{align}
where $m^2_{\tilde{\nu}_X}$ is the sneutrino mass-squared, $m^2_{\tilde{l}_X}$ is the slepton mass-squared and $M_{\tilde{\chi}^{0,-}_A}$ is the neutralino (chargino) mass. The loop integrals $I_{1,2}$ and $J_{1,2}$ are defined by
\begin{align}
I_1(r) &= \frac{1}{12(1-r)^4}(2+3r-6r^2+r^3 + 6r\log r)\\
I_2(r) &= \frac{1}{2(1-r)^3}(-3+4r-r^2 -2\log r) \\
J_1(r) &= \frac{1}{12(1-r)^4}(1-6r+3r^2+2r^3 - 6r^2\log r)\\
J_2(r) &= \frac{1}{2(1-r)^3}(1-r^2 + 2\log r)
\end{align}
and the matrices $C$ and $N$ are defined as
\begin{align}
C_{iAX}^R &= -g_2 (O_R)_{A1} U^\nu_{X,i}\\
C_{iAX}^L &= g_2 \frac{m_{l_i}}{\sqrt{2}M_W\cos\beta}(O_L)_{A2} U^\nu_{X,i}
\end{align}
where $A=1,2$ and $X=1,2,3$, 
\begin{align}
N_{iAX}^R &= -\frac{g_2}{\sqrt{2}}\left\{ \left[ -(O_N)_{A2} - (O_n)_{A1} \tan \theta_W\right] U^l_{X,i} + \frac{m_{l_i}}{M_W \cos\beta}(O_N)_{A3} U^l_{X,i+3}\right\}\\
N_{iAX}^L &= -\frac{g_2}{\sqrt{2}}\left\{ \frac{m_{l_i}}{M_W\cos\beta} (O_N)_{A3} U^l_{X,i} + 2(O_N)_{A1} \tan\theta_W U^l_{X,i+3}\right\}
\end{align}
where $A=1,\ldots, 4$ and $X=1,\ldots, 6$. The matrices $(O_N)$ are those acting on the neutralino mass matrix such that $O_N M_N O_N^T$ is diagonal.

\subsection{Expressions for $\mu \to e$ conversion in Nuclei}

The effective Hamiltonian for the interactions of the muon and electron with the vector quark current can be written as
\beq
\mathcal{H} = g_{LV}^q(\bar{e}\gamma_\nu P_L\mu)(\bar{q}\gamma^\nu q) + g_{RV}^q(\bar{e}\gamma_\nu P_R\mu)(\bar{q}\gamma^\nu q)
\eeq

with $g_{L(R)V}^q$ being Wilson coefficients. These can be computed in the full eigenstate basis, but we give the expressions for them here in the mass insertion approximation. It was verified numerically that the full eigenstate basis calculation matched the mass insertion approximation.
\beq
g_{L(R)V}^q = g_{L(R)V}^{q,box} + g_{L(R)V}^{q,\gamma} + g_{L(R)V}^{q, Z}
\eeq
where in the limit where the Wino mass is much less than the scalar masses, $|M_{\tilde{W}}| \ll m_{\tilde{l},\tilde{q}}$, the leading contributions are given by \cite{Altmannshofer:2013lfa}
\beq
g_{LV}^{q,box} = \frac{g_2^4}{(4\pi)^2 m_{\tilde{q}}^2}\left(\frac{\Delta m_{\tilde{l}}^2}{m_{\tilde{l}}^2}\right)\frac{5}{4} f\left(\frac{m_{\tilde{l}}^2}{m_{\tilde{q}}^2}\right),
\eeq

\beq
g_{LV}^{u,\gamma-peng.}=-2g_{LV}^{d,\gamma-peng.} = \frac{-2 e^2 g_2^2}{(4\pi)^2 m_{\tilde{l}}^2}\left(\frac{\Delta m_{\tilde{l}}^2}{m_{\tilde{l}}^2}\right)\left\{ \frac{1}{4} + \frac{1}{9} \log\left(\frac{|M_{\tilde{W}}|^2}{m_{\tilde{l}}^2}\right)\right\},
\eeq

\begin{align}
\nonumber g_{LV}^{u,Z-peng.}&=\frac{-\left(1-\frac{4}{3}\sin^2\theta_W\right)}{\left(1-\frac{8}{3}\sin^2\theta_W\right)}g_{LV}^{d,Z-peng.} \\ \nonumber&= \frac{-g_2^4}{(4\pi)^2 m_{\tilde{l}}^2}\left(\frac{\Delta m_{\tilde{l}}^2}{m_{\tilde{l}}^2}\right)\frac{1}{16}\left(1-\frac{8}{3}\sin^2\theta_W\right)\\&\times \left\{\cos\beta f_1\left(\frac{|M_{\tilde{W}}|^2}{m_{\tilde{l}}^2},\frac{|\mu|^2}{m_{\tilde{l}}^2}\right) + \sin\beta f_2\left(\frac{|M_{\tilde{W}}|^2}{m_{\tilde{l}}^2},\frac{|\mu|^2}{m_{\tilde{l}}^2}\right) + \frac{\mu M_{\tilde{W}}}{m_{\tilde{l}}^2}f_3\left(\frac{|M_{\tilde{W}}|^2}{m_{\tilde{l}}^2},\frac{|\mu|^2}{m_{\tilde{l}}^2}\right) \right\}
\end{align}
with $\Delta m_{\tilde{l}}^2 = M_{3/2}^2 (0.9c \times\epsilon)$ and $m_{\tilde{l}}^2 = M_{3/2}^2(1-0.9c)$, where $0.9c \simeq 0.5$. We reproduce here the loop functions given in Appendix B of \cite{Altmannshofer:2013lfa} for convenience:
\begin{align}
f(x) &= \frac{1}{8(1-x)} + \frac{x\log x}{8(1-x)^2},\\
\nonumber f_1(x,y) &= \frac{x^3(3-9y) + (y-3)y^2 + x^2(3y-1)(1+4y) + xy(y(13-11y)-4)}{2(1-x)^2(1-y)^2(x-y)^2}\\ \nonumber&+ \frac{x(2x^3 + 2y^2 + 3xy(1+y) -x^2(1+9y))}{(1-x)^3(x-y)^3}\log x \\&+\frac{y^2(y+x(7y-5)-3x^2)}{(1-y)^3(x-y)^3}\log y,\\
f_2(x,y) &= \frac{x^3(1-3y) + 3(y-3)y^2 + x(y-3)y(y+4) + x^2(y(13-4y)-11)}{2(1-x)^2(1-y)^2(x-y)^2}\\ \nonumber&+ \frac{x(2x^3 + 2y^2 + 3x^2(1+y) -xy(9+y))}{(1-x)^3(y-x)^3}\log x \\&+\frac{y^2(x^2+x(7-5y)-3y)}{(1-y)^3(y-x)^3}\log y,\\
\nonumber f_3(x,y) &= -\frac{12(x+y+x^2+y^2+x^2y+y^2x-6xy}{(1-x)^2(1-y)^2(x-y)^2}\\ \nonumber &+\frac{24x(x^2-y)}{(1-x)^3(y-x)^3}\log x + \frac{24y(y^2-x)}{(1-y)^3(x-y)^3}\log y ,
\end{align}
These loop functions take into account the separation of scale between the gauginos, the $\mu-$term and the scalar masses.

The overlap integrals calculated in \cite{Kitano:2002mt} are given here for $_{13}^{27}$Al:
\begin{itemize}
\item $D=0.0357 (m_\mu)^{5/2}$,
\item $V^{(p)} = 0.0159 (m_\mu)^{5/2}$,
\item $V^{(n)} = 0.0169 (m_\mu)^{5/2}$.
\end{itemize}


\begin{thebibliography}{20}

\bibitem{Borzumati:1986qx}
  F.~Borzumati and A.~Masiero,
  Phys.\ Rev.\ Lett.\  {\bf 57} (1986) 961.


  
\bibitem{Raidal:2008jk}
  M.~Raidal, A.~van der Schaaf, I.~Bigi, M.~L.~Mangano, Y.~K.~Semertzidis, S.~Abel, S.~Albino and S.~Antusch {\it et al.},
  Eur.\ Phys.\ J.\ C {\bf 57} (2008) 13
  [arXiv:0801.1826 [hep-ph]].
  
\bibitem{Arganda:2005ji}
  E.~Arganda and M.~J.~Herrero,
  Phys.\ Rev.\ D {\bf 73} (2006) 055003
  [hep-ph/0510405].
  
\bibitem{Gabbiani:1996hi}
  F.~Gabbiani, E.~Gabrielli, A.~Masiero and L.~Silvestrini,
  Nucl.\ Phys.\ B {\bf 477} (1996) 321
  [hep-ph/9604387].
  

\bibitem{Universality}

   J.~R.~Ellis and D.~V.~Nanopoulos,
  Phys.\ Lett.\ B {\bf 110} (1982) 44.

R.~Barbieri and R.~Gatto,
  Phys.\ Lett.\ B {\bf 110} (1982) 211.
  
  
  M.~Dine, A.~Kagan and S.~Samuel,
  Phys.\ Lett.\ B {\bf 243} (1990) 250.
  
  L.~E.~Ibanez and D.~Lust,
  Nucl.\ Phys.\ B {\bf 382} (1992) 305
  [hep-th/9202046].
  
  V.~S.~Kaplunovsky and J.~Louis,
  Phys.\ Lett.\ B {\bf 306} (1993) 269
  [hep-th/9303040].
  
   R.~Barbieri, J.~Louis and M.~Moretti,
  Phys.\ Lett.\ B {\bf 312} (1993) 451
   [Erratum-ibid.\ B {\bf 316} (1993) 632]
  [hep-ph/9305262].
  
   B.~de Carlos, J.~A.~Casas and C.~Munoz,
  Phys.\ Lett.\ B {\bf 299} (1993) 234
  [hep-ph/9211266].
  
   B.~de Carlos, J.~A.~Casas and C.~Munoz,
  Nucl.\ Phys.\ B {\bf 399} (1993) 623
  [hep-th/9204012].
  
   A.~Brignole, L.~E.~Ibanez and C.~Munoz,
  Nucl.\ Phys.\ B {\bf 422} (1994) 125
   [Erratum-ibid.\ B {\bf 436} (1995) 747]
  [hep-ph/9308271].
  
    D.~Matalliotakis and H.~P.~Nilles,
  Nucl.\ Phys.\ B {\bf 435} (1995) 115
  [hep-ph/9407251].
  
  D.~Choudhury, F.~Eberlein, A.~Konig, J.~Louis and S.~Pokorski,
  Phys.\ Lett.\ B {\bf 342} (1995) 180
  [hep-ph/9408275].
  
  
\bibitem{Acharya:2007rc}
  B.~S.~Acharya, K.~Bobkov, G.~L.~Kane, P.~Kumar and J.~Shao,
  Phys.\ Rev.\ D {\bf 76} (2007) 126010
  [hep-th/0701034].

\bibitem{Acharya:2008hi}
  B.~S.~Acharya and K.~Bobkov,
  JHEP {\bf 1009} (2010) 001
  [arXiv:0810.3285 [hep-th]].
  
\bibitem{Acharya:2008zi}
  B.~S.~Acharya, K.~Bobkov, G.~L.~Kane, J.~Shao and P.~Kumar,
  Phys.\ Rev.\ D {\bf 78} (2008) 065038
  [arXiv:0801.0478 [hep-ph]].
  
\bibitem{Acharya:2011te}
  B.~S.~Acharya, G.~Kane, E.~Kuflik and R.~Lu,
  JHEP {\bf 1105} (2011) 033
  [arXiv:1102.0556 [hep-ph]].
  
  
\bibitem{Acharya:2012tw}
  B.~S.~Acharya, G.~Kane and P.~Kumar,
  Int.\ J.\ Mod.\ Phys.\ A {\bf 27} (2012) 1230012
  [arXiv:1204.2795 [hep-ph]].
  

  
  
  
  
\bibitem{Hisano:1995cp}
  J.~Hisano, T.~Moroi, K.~Tobe and M.~Yamaguchi,
  Phys.\ Rev.\ D {\bf 53} (1996) 2442
  [hep-ph/9510309].
  
\bibitem{Casas:2001sr}
  J.~A.~Casas and A.~Ibarra,
  Nucl.\ Phys.\ B {\bf 618} (2001) 171
  [hep-ph/0103065].
  
\bibitem{Ellis:2001xt}
  J.~R.~Ellis, J.~Hisano, S.~Lola and M.~Raidal,
  Nucl.\ Phys.\ B {\bf 621} (2002) 208
  [hep-ph/0109125].
  
  J.~R.~Ellis, J.~Hisano, M.~Raidal and Y.~Shimizu,
  Phys.\ Rev.\ D {\bf 66} (2002) 115013
  [hep-ph/0206110].
  
  J.~R.~Ellis and M.~Raidal,
  Nucl.\ Phys.\ B {\bf 643} (2002) 229
  [hep-ph/0206174].
  
  
  
\bibitem{Moroi:2013sfa}
  T.~Moroi and M.~Nagai,
  Phys.\ Lett.\ B {\bf 723} (2013) 107
  [arXiv:1303.0668 [hep-ph]].

  
\bibitem{Kane:2009kv}
  G.~Kane, P.~Kumar and J.~Shao,
  Phys.\ Rev.\ D {\bf 82} (2010) 055005
  [arXiv:0905.2986 [hep-ph]].
  
\bibitem{Ellis:2014tea}
  S.~A.~R.~Ellis and G.~L.~Kane,
  arXiv:1405.7719 [hep-ph].


\bibitem{Adam:2013mnn}
  J.~Adam {\it et al.}  [MEG Collaboration],
  Phys.\ Rev.\ Lett.\  {\bf 110} (2013) 201801
  [arXiv:1303.0754 [hep-ex]].

\bibitem{Baldini:2013ke}
  A.~M.~Baldini, F.~Cei, C.~Cerri, S.~Dussoni, L.~Galli, M.~Grassi, D.~Nicolo and F.~Raffaelli {\it et al.},
  arXiv:1301.7225 [physics.ins-det].
  
\bibitem{Bellgardt:1987du}
  U.~Bellgardt {\it et al.}  [SINDRUM Collaboration],
  Nucl.\ Phys.\ B {\bf 299} (1988) 1.
  
\bibitem{Blondel:2013ia}
  A.~Blondel, A.~Bravar, M.~Pohl, S.~Bachmann, N.~Berger, M.~Kiehn, A.~Schoning and D.~Wiedner {\it et al.},
  arXiv:1301.6113 [physics.ins-det].
  
\bibitem{Aubert:2009ag}
  B.~Aubert {\it et al.}  [BaBar Collaboration],
  Phys.\ Rev.\ Lett.\  {\bf 104} (2010) 021802
  [arXiv:0908.2381 [hep-ex]].
  
\bibitem{Hayasaka:2013dsa}
  K.~Hayasaka [Belle and Belle II Collaboration],
  J.\ Phys.\ Conf.\ Ser.\  {\bf 408} (2013) 012069.
  
\bibitem{Bertl:2006up}
  W.~H.~Bertl {\it et al.}  [SINDRUM II Collaboration],
  Eur.\ Phys.\ J.\ C {\bf 47} (2006) 337.
  
\bibitem{Kaulard:1998rb}
  J.~Kaulard {\it et al.}  [SINDRUM II Collaboration],
  Phys.\ Lett.\ B {\bf 422} (1998) 334.
  
 \bibitem{Comet}
 COMET Collaboration, http://comet.kek.jp/Documents.html
  
\bibitem{Abrams:2012er}
  R.~J.~Abrams {\it et al.}  [Mu2e Collaboration],
  arXiv:1211.7019 [physics.ins-det].
  
\bibitem{Paradisi:2005fk}
  P.~Paradisi,
  JHEP {\bf 0510} (2005) 006
  [hep-ph/0505046].
  
    
\bibitem{Kuno:1999jp}
  Y.~Kuno and Y.~Okada,
  Rev.\ Mod.\ Phys.\  {\bf 73} (2001) 151
  [hep-ph/9909265].
  
  
\bibitem{Altmannshofer:2013lfa}
  W.~Altmannshofer, R.~Harnik and J.~Zupan,
  JHEP {\bf 1311} (2013) 202
  [arXiv:1308.3653 [hep-ph]].
  
\bibitem{Kitano:2002mt}
  R.~Kitano, M.~Koike and Y.~Okada,
  Phys.\ Rev.\ D {\bf 66} (2002) 096002
   [Phys.\ Rev.\ D {\bf 76} (2007) 059902]
  [hep-ph/0203110].
  
  
  
\bibitem{Acharya:2004qe}
  B.~S.~Acharya and S.~Gukov,
  Phys.\ Rept.\  {\bf 392} (2004) 121
  [hep-th/0409191].
  
\bibitem{Bourjaily:2009ci}
  J.~L.~Bourjaily,
  arXiv:0905.0142 [hep-th].
  
  
\bibitem{Brignole:1997dp}
  A.~Brignole, L.~E.~Ibanez and C.~Munoz,
  Adv.\ Ser.\ Direct.\ High Energy Phys.\  {\bf 21} (2010) 244
  [hep-ph/9707209].

  
  
\bibitem{Ramond:1993kv}
  P.~Ramond, R.~G.~Roberts and G.~G.~Ross,
  Nucl.\ Phys.\ B {\bf 406} (1993) 19
  [hep-ph/9303320].
 
  
  
  
\bibitem{Kane:2011kj}
  G.~Kane, P.~Kumar, R.~Lu and B.~Zheng,
  Phys.\ Rev.\ D {\bf 85} (2012) 075026
  [arXiv:1112.1059 [hep-ph]].
  
  \bibitem{Kane:2012qr}
  G.~Kane, R.~Lu and B.~Zheng,
  Int.\ J.\ Mod.\ Phys.\ A {\bf 28} (2013) 1330002
  [arXiv:1211.2231 [hep-ph]].
  
\bibitem{Ellis:2014kla}
  S.~A.~R.~Ellis, G.~L.~Kane and B.~Zheng,
  arXiv:1408.1961 [hep-ph].
  
  
\bibitem{CMS:2014xfa}
  V.~Khachatryan {\it et al.}  [CMS and LHCb Collaborations],
  Nature (2015)
  [arXiv:1411.4413 [hep-ex]].
  
  
\bibitem{Martin:1993zk}
  S.~P.~Martin and M.~T.~Vaughn,
  Phys.\ Rev.\ D {\bf 50} (1994) 2282
   [Phys.\ Rev.\ D {\bf 78} (2008) 039903]
  [hep-ph/9311340].
  
\bibitem{Allanach:2001kg}
  B.~C.~Allanach,
  Comput.\ Phys.\ Commun.\  {\bf 143} (2002) 305
  [hep-ph/0104145].
  
  
 
  
\bibitem{Hisano:1998cx}
  J.~Hisano, D.~Nomura, Y.~Okada, Y.~Shimizu and M.~Tanaka,
  Phys.\ Rev.\ D {\bf 58} (1998) 116010
  [hep-ph/9805367].
  
\bibitem{Calibbi:2015kja}
  L.~Calibbi, I.~Galon, A.~Masiero, P.~Paradisi and Y.~Shadmi,
  arXiv:1502.07753 [hep-ph].
  
\bibitem{Witten:2001bf}
  E.~Witten,
  hep-ph/0201018.
  


  
  
\bibitem{Agashe:2014kda}
  K.~A.~Olive {\it et al.}  [Particle Data Group Collaboration],
  Chin.\ Phys.\ C {\bf 38} (2014) 090001.
  
  

  
  
  
  
  
\bibitem{Bagger:1997gg}
  J.~A.~Bagger, K.~T.~Matchev and R.~J.~Zhang,
  Phys.\ Lett.\ B {\bf 412} (1997) 77
  [hep-ph/9707225].
  
\bibitem{Ciuchini:1998ix}
  M.~Ciuchini, V.~Lubicz, L.~Conti, A.~Vladikas, A.~Donini, E.~Franco, G.~Martinelli and I.~Scimemi {\it et al.},
  JHEP {\bf 9810} (1998) 008
  [hep-ph/9808328].
  
  

  

  

  

  
  


  


\end{thebibliography}
\end{document}